\documentclass[11pt,oneside,english, reqno]{amsart}
\usepackage{mathpazo}
\usepackage[scaled=0.92]{helvet}
\usepackage{courier}
\usepackage[T1]{fontenc}
\usepackage[latin9]{inputenc}
\usepackage[a4paper]{geometry}
\usepackage{amsthm}
\usepackage{amssymb}
\usepackage{setspace}
\usepackage[authoryear]{natbib}
\usepackage[ruled,linesnumbered]{algorithm2e}
\usepackage[hidelinks]{hyperref}
\usepackage{graphicx}
\usepackage{subfigure}
\numberwithin{equation}{section}

\usepackage{soul}
\usepackage{bbm}
\usepackage{caption} 
\makeatletter

\theoremstyle{plain}

\numberwithin{equation}{section}
\renewcommand{\v}[1]{\boldsymbol{#1}}
\makeatother
\usepackage{amsmath}
\usepackage{ifthen}

\newcommand{\bb}[1]{\boldsymbol{#1}}
\newcommand{\btheta}{\bb{\theta}}

\usepackage{amsfonts}
\usepackage{mathtools}
\usepackage{xcolor,color}

\begin{document}
\title[Dynamic linear regression forecasting]{Dynamic linear regression models for forecasting time series with semi-long memory errors}
\author{Thomas Goodwin$^{\,\dagger}$, Matias Quiroz$^{\,\ddagger\,\dagger}$ and Robert Kohn$^{\,\S\,\P}$}
\thanks{$\dagger$: \textit{Human Technology Institute, University of Technology Sydney}.\\ $\ddagger$: \textit{School of Mathematical and Physical Sciences, University of Technology Sydney}.\\
$\S$: \textit{School of Economics, University of New South Wales}. 
\\ $\P$: \textit{Data Analytics Center for Resources and Environments (DARE)}}

\begin{abstract}
Dynamic linear regression models forecast the values of a time series based on a linear combination of a set of exogenous time series while incorporating a time series process for the error term. This error process is often assumed to follow a stationary autoregressive integrated moving average (ARIMA) model, or its seasonal variants, which are unable to capture a long-range dependence structure (long memory) of the error process. We propose a novel dynamic linear regression model that incorporates the long-range dependence feature of the errors and show that the proposed error process may: (i) have a significant impact on the posterior uncertainty of the estimated regression parameters and (ii) improve the model's forecasting ability. We develop a Markov chain Monte Carlo method to fit general dynamic linear regression models based on a frequency domain approach that enables fast, asymptotically exact Bayesian inference for large datasets. We demonstrate that our approximate algorithm is faster than the traditional time domain approaches, such as the Kalman filter and the multivariate Gaussian likelihood, while producing a highly accurate approximation to the posterior. The method is illustrated in simulated examples and two energy forecasting applications, showing that it outperforms approaches that do not account for semi-long memory, as well as a state-of-the-art neural-network-based forecasting procedure.
\end{abstract}

\maketitle

\subsection*{Acknowledgements}
The authors used an AI language model (ChatGPT, OpenAI) to improve grammar and flow. All scientific content and interpretations are the authors' own. 

TG and MQ are partially supported by the Marine Ecosystems Research Mobilising AI and Data (MERMAID) Collaboration (CLB-3127).

\newpage
\section{Introduction} \label{section: DLR introduction}

Forecasting time series data plays an important role in various fields, such as engineering, economics, and climate sciences. Forecasting a single output time series may be more accurate by using linear combinations of exogenous variables that explain some of its historical variation \citep{hyndman2018forecasting}. However, the linear combination of the exogenous time series often does not capture all the serial correlation present in the output time series, resulting in autocorrelated errors. Dynamic linear regression models \citep{pankratz2012forecasting} provide a framework that relates the output time series to a linear combination of the exogenous time series and models the resulting error term as a time series process to account for serially correlated errors. Dynamic linear regression (DLR) models are a particular case of the more general class of transfer function models \citep{box2015time}.

The standard dynamic linear regression model assumes that the error process follows an autoregressive integrated moving average process (ARIMA). This approach allows the error process to have an autoregressive component and a moving average component for its white noise error term. While this error process accommodates a wide range of stationary processes, it requires a large number of autoregressive components to accommodate error processes that show significant autocorrelation between distant time periods. To capture such autocorrelations of the error process parsimoniously, we employ the autoregressive tempered fractionally integrated moving average (ARTFIMA) \citep{meerschaert2014tempered, sabzikar2019parameter}. The ARTFIMA class nests the well-known autoregressive fractionally integrated moving average (ARFIMA) \citep{granger1980introduction} model, which is useful for time series with so-called long memory. There is a large literature on long-memory processes and their application in regression settings. Section \ref{subsec:related_work} provides a review of those most relevant to our problem.



The ARTFIMA model has several closely related advantages over the ARFIMA model, which are particularly relevant for the frequency domain estimation approach proposed in this paper. The key distinction is the type of dependence structure implied by the two models. In an ARFIMA process, the autocovariance function decays slowly according to a power law and is not absolutely summable \citep{granger1980introduction,mcleod1978preservation}. Such processes are referred to as having long memory. In contrast, the ARTFIMA model exhibits long-range dependence over a range of lags, but its autocovariance function eventually decays exponentially fast and is therefore absolutely summable. This behaviour is referred to as semi-long memory and is typically easier to analyse \citep{sabzikar2019parameter}. This difference in dependence structure has important implications in the spectral domain. For ARFIMA processes, the spectral density diverges as the frequency approaches zero, whereas for ARTFIMA processes, it remains bounded. Empirically, a stylised fact of estimated power spectra is that they are bounded at low frequencies, and ARTFIMA models have been shown to capture this behaviour better than ARFIMA models \citep{meerschaert2014tempered,sabzikar2019parameter}. This feature is particularly important in our setting, since we develop an estimation method based on a parametric Whittle likelihood \citep{whittle1953estimation}, which approximates the time-domain likelihood by representing the information in the data through the spectral density. For long-memory processes, the Whittle approximation is known to perform poorly, especially at small frequencies \citep{robinson1995log,Rousseau2012bayesnonparametric}. Section~\ref{sec: DLR periodogram simulations} illustrates this issue for dynamic linear regression models with ARFIMA errors and shows that it is mitigated when using ARTFIMA errors. This does not imply that likelihood-based inference for long-memory models is infeasible; rather, the issue concerns the accuracy of the Whittle approximation to the sampling distribution of the full time-domain likelihood, particularly at low frequencies. We also show that resorting to the time-domain likelihood for DLR models with ARFIMA errors \citep{doornik2003computational} is computationally much more costly than our frequency domain approach, while offering no systematic gains in predictive accuracy.

The parameters in dynamic linear regression models consist of the regression coefficients that form the linear combination of the exogenous time series and the parameters of the error process. For the standard dynamic linear regression models with ARIMA errors that are normally distributed, efficient likelihood-based inference can be carried out by finding the finite-dimensional state space representation of the model and using the Kalman filter to integrate out the unobserved time-varying error terms \citep{anderson2005optimal}. This may still be computationally costly with many time observations, especially in Bayesian inference, which typically requires many posterior samples for reliable inference. For each such sample, the Kalman filter needs to cycle through all the observations, which can be prohibitively expensive, especially for large time series. The Whittle log-likelihood can be derived using large sample properties of the so-called periodogram data, which are formed via the discrete Fourier transform of the time domain data. The Whittle log-likelihood is directly a function of the regression coefficients and the error process parameters without needing to integrate out unobserved error terms. However, in DLR models, the periodogram data becomes a function of the regression coefficients, which requires recomputing the discrete Fourier transform in every iteration. We show how this can be avoided, and thus our algorithm requires computing the discrete Fourier transform only once before running the sampling algorithm, thereby obtaining significant computational gains compared to the time domain log-likelihood based on the Kalman filter. Our frequency domain approach also applies to a dynamic linear regression model with ARTFIMA errors, where a finite-dimensional state space representation is not readily available.

To summarise, our article makes two important contributions. First, we propose a frequency domain estimation approach for dynamic linear models that significantly outperforms estimation approaches based on the time domain likelihood in terms of computing time, especially when a finite-dimensional state space representation of the model is unavailable. Second, we utilise a semi-long memory process for the error process and show that it may provide more accurate forecasts than both the standard dynamic linear model and models with long-memory errors, particularly for datasets with complex long-range dependence, as well as a popular neural-network-based forecasting approach. Moreover, we demonstrate that the semi-long memory process may significantly impact the posterior distribution of the regression parameters. 

The rest of the article is organised as follows. Section \ref{section: dynamic linear models} reviews existing dynamic linear models and presents our extension. Section \ref{section: DLR methodology} introduces the necessary frequency domain tools, outlines our estimation method, and validates its performance relative to the time domain likelihood. Section \ref{section: DLR simulated example} includes relevant simulation studies. Section \ref{section: DLR applications} presents applications for two real-world electricity demand datasets.

\section{Dynamic linear regression models}\label{section: dynamic linear models}

\subsection{Standard dynamic linear regression models}\label{subsection: ARIMA error}
Let $\v X_t = (X_{1t},\dots ,X_{mt})^\top \in \mathbb{R}^m$ be a set of $m$ exogenous stationary time series observed at time $t$. A dynamic linear regression models the scalar output time series $Y_t \in \mathbb{R}$ as a linear combination of the exogenous $\v X_t$, i.e.\
 \begin{align}
Y_t & = \v X_t^\top \v \beta + \eta_t,  \label{eq:DLR}
\end{align}
where $\v \beta \in  \mathbb{R}^m$ is a vector of regression coefficients and $\eta_t \in  \mathbb{R}$ is a zero-mean error process. The error process contains all unobserved factors that affect $Y_t$, and we assume that $\mathrm{E}(\eta_t |  \v X_t)=0$, i.e. it is uncorrelated with each of the elements in $\v X_t$ and hence $\v X_t$ is exogenous. 

When the error process $\eta_t$ is an independent series, \eqref{eq:DLR} is a standard linear regression (if it has a constant variance) and can easily be estimated using standard linear regression approaches (or modified versions thereof if heteroscedasticity is present). However, in many applications, the unobserved factors are time-varying, resulting in serially correlated errors $\eta_t$. The standard dynamic linear regression model assumes that $\eta_t$ is an autoregressive integrated moving average process, denoted ARIMA$(p,d,q)$ and defined as
\begin{align}
\phi_p(B)\Delta^d \eta_t & = \psi_q(B)\varepsilon_t, \label{eq:Standard_DLR}
\end{align}
where  $\phi_p(B) = 1-\sum^{p}_{i=1}\phi_iB^i$ and $\psi_q(B) = 1+\sum^{q}_{i=1}\psi_iB^i$ are the autoregressive and moving average lag polynomials, respectively, with the lag operator $B$ such that $B^i \eta_t = \eta_{t-i}$. The differencing operator $\Delta^d$, for $d=0, 1, 2,\dots$, is defined as $\Delta^d \eta_t = (1 - B)^d\eta_t$. Finally, $\varepsilon_t$ is the white noise error. When the error process $\eta_t$ exhibits seasonality with seasonal period $s$, this can be modelled by a seasonal ARIMA process, denoted ARIMA($p,d,q)(P,D,Q)_s$ and defined as
\begin{align}
\phi_p(B)\phi_P^\star(B^s)\Delta^d \Delta^D_s\eta_t = \psi_q(B)\psi_Q^\star(B^s)\varepsilon_t , \label{eq:Standard_DLR_seasonal}
\end{align}
where $\phi_P^\star(B^s) = 1-\sum^{P}_{i=1}\phi^\star_iB^{is}$ and $\psi_Q^\star(B^s) = 1+\sum^{Q}_{i=1}\psi^\star_iB^{is}$ are the seasonal autoregressive and seasonal moving average lag polynomials, and $\Delta^D_s = (1-B^s)^D$, for $D=0, 1, 2,\dots$, is the seasonal differencing operator \citep{box2015time}. 

To carry out inference in \eqref{eq:DLR}, it is typically assumed that $\eta_t$ is normally distributed. The resulting log-likelihood is then a multivariate Gaussian distribution with a Toeplitz covariance matrix given that $\eta_t$, or after integer differencing of $\eta_t$, is stationary \citep{doornik2003computational}. Inversion of such a matrix is usually performed via the Levinson-Durbin algorithm \citep{levinson1946wiener, durbin1960fitting} in $\mathcal{O}(T^2)$ operations. However, recent approaches, known as superfast Toeplitz algorithms, can solve the matrix system in $\mathcal{O}(T\log_2^2(T))$ operations, which becomes more efficient than the Levinson-Durbin algorithm when $T>256$ \citep{ammar1988superfast}. A method that scales even better is to find the finite-dimensional state space representation of the model in \eqref{eq:DLR} with the error process following either \eqref{eq:Standard_DLR} or \eqref{eq:Standard_DLR_seasonal}, in which the resulting likelihood can be evaluated in $\mathcal{O}(T)$ using the Kalman filter.

\subsection{Long memory dynamic linear regression models} \label{subsection: ARFIMA error}

In many applications, a pair of observations separated by a long time interval exhibit a non-negligible correlation. The standard dynamic linear regression model with the error process in\eqref{eq:Standard_DLR} has an autocovariance function whose absolute value decays exponentially fast and thus cannot capture this feature. A common approach to model time series with long memory is to consider so-called fractional differencing. 

In a dynamic linear regression setting, \cite{doornik2004inference} propose to model the error process $\eta_t$ in \eqref{eq:DLR} with an autoregressive fractionally integrated moving average process \citep{granger1980introduction, hosking1981frac}, denoted ARFIMA$(p, d, q)$. This model is, for $d\notin \mathbb{Z}$,
\begin{align}
\phi_p(B)\Delta^d \eta_t & = \psi_q(B)\varepsilon_t, \label{eq:ARFIMA_DLR}
\end{align}
where $\phi_p(B)$ and $\psi_q(B)$ are defined as in \eqref{eq:Standard_DLR}. The fractional differencing operator is defined via the fractional binomial theorem
\begin{align*}
 \Delta^d \eta_t & = (1 - B)^d\eta_t \\
 & = \sum_{j=0}^{\infty} {\binom{d}{j}} (-B)^j \eta_t \\
 & = \sum_{j=0}^{\infty} (-1)^j \frac{\Gamma(1+d)}{\Gamma(1+d-j)j!}  \eta_{t-j},
\end{align*}
where $\Gamma(a)=\int \exp(-t)t^{a-1}dt$ using $a!=\Gamma(1+a)$. Provided that the roots of $\phi_p(z)$ are outside the unit circle, the ARFIMA process is stationary if $-0.5 < d < 0.5$ and has long memory when $0 < d < 0.5$ \citep{granger1980introduction}. \cite{doornik2004inference} also propose a seasonal extension by modelling $\eta_t$ as a seasonal ARFIMA($p,d,q)(P,D,Q)_s$ with $d \notin \mathbb{Z}$ and integer $D$. A possible extension of this model is to allow for non-integer $D$ as in \cite{bisognin2009}. 

\cite{doornik2003computational} show that maximum likelihood estimation for ARFIMA models is possible in $\mathcal{O}(T^2)$ time, and this also applies for the model in \eqref{eq:DLR} with the error process following \eqref{eq:ARFIMA_DLR}. \cite{chan1998state} show that there is no finite-dimensional state space representation of an ARFIMA process. They propose an infinite-dimensional state space representation for which the Kalman filter can be computed in $T$ steps; however, the resulting computation is $\mathcal{O}(T^3)$.

\subsection{Related work on long-memory processes}\label{subsec:related_work}
Long-memory processes have been studied extensively for several decades, and there is a rich literature on frequentist estimation of ARFIMA models; see, for example, \citet{Estimation_ARFIMA_GEWEKE}, \citet{MLE_ARFIMA_SOWELL}, \citet{Approximate_ARFIMA_MLE}, and \citet{MLE_ARFIMA_Whittle_taper}. Introductory treatments and broader reviews are provided in \citet{granger1980introduction} and \citet{Statistics_for_long_memory_processes}. Much of this literature establishes frequentist and Whittle-based estimation methods for long-memory processes, including regression models with long-memory errors and more general locally stationary frameworks; see, for example, \citet{palma2010efficient}.

Our focus is instead on full Bayesian inference using a frequency domain likelihood in which the periodogram ordinates are treated as independent observations. In this setting, correct specification of the sampling distribution at low frequencies is crucial, since each frequency contributes directly to the likelihood and hence to posterior uncertainty quantification. As shown in Section~\ref{sec: DLR periodogram simulations}, for dynamic linear regression models with ARFIMA errors, the scaled low-frequency periodogram ordinates do not follow their assumed exponential distribution, which undermines the use of a Whittle likelihood as a full Bayesian likelihood.

In contrast, as shown below, the ARTFIMA model introduces a tempering parameter that preserves long-range dependence over intermediate lags while ensuring that the spectral density remains bounded at zero frequency. This semi-long memory property restores the validity of the Whittle likelihood as a full likelihood approximation in large samples, making it well-suited for Bayesian inference in dynamic linear regression models. While ARFIMA errors have been used in DLR applications such as inflation \citep{doornik2004inference,DLR_ARFIMA_inflation_Angola}, steel consumption \citep{DLR_ARFIMA_steel_consumption}, and tourism demand \citep{DLR_ARFIMA_Japan_unemployment}, the use of ARTFIMA errors in a Bayesian DLR framework has, to our knowledge, not previously been considered.

\subsection{Semi-long memory dynamic linear regression models}\label{subsection: ARTFIMA error}

The autoregressive tempered fractional integrated moving-average \citep{meerschaert2014tempered, sabzikar2019parameter} (ARTFIMA) extends the ARFIMA model, incorporating a tempering parameter $\lambda$. The model for the error process is then 
\begin{align}
\phi_p(B)\Delta^{d,\lambda} \eta_t & = \psi_q(B)\varepsilon_t, \label{eq:ARTFIMA_DLR}
\end{align}
where $\phi_p(B)$ and $\psi_q(B)$ are defined as in \eqref{eq:Standard_DLR} and for $d \notin \mathbb{Z}$ and $\lambda > 0$ the tempered fractional differencing operator is
\begin{align}
\label{eq:Tempered_frac_diff}
 \Delta^{d,\lambda} \eta_t & = (1 - \exp(-\lambda)  B)^d\eta_t \nonumber\\
 & = \sum_{j=0}^{\infty} {\binom{d}{j}} (-B)^j \eta_t \nonumber \\
 & = \sum_{j=0}^{\infty} (-1)^j \frac{\Gamma(1+d)}{\Gamma(1+d-j)j!}\exp(-\lambda j)  \eta_{t-j}.
\end{align}
Note that when $\lambda = 0$, the ARTFIMA model is a stationary ARFIMA model if $-0.5 < d < 0.5$, provided the roots of the polynomial $\phi_p(z)$ are outside the unit circle. When $\lambda = 0$ and $d$ is an integer, the ARTFIMA model becomes an ARIMA model. The ARTFIMA model is stationary for all $d \notin \mathbb{Z}$ and $\lambda > 0$ provided that the root condition of the polynomial $\phi_p(z)$  is satisfied \citep{sabzikar2019parameter}. Section \ref{section: DLR introduction} outlines the advantages the ARTFIMA process has over ARFIMA. 

Similarly to the ARFIMA model, a finite-dimensional state space representation that allows efficient Kalman filter computations is unavailable for the ARTFIMA model. Instead, the log-likelihood can be computed using a multivariate Gaussian distribution with a Toeplitz covariance matrix formed via the autocovariance function in \citet[Theorem 2.5 (b)]{sabzikar2019parameter}. However, Section \ref{section: DLR methodology} describes a more computationally efficient approach based on a frequency domain log-likelihood for the periodogram data. The approach relies on the spectral density of the process, which we now describe.

\begin{figure}[t] 
    \centering
    \includegraphics[width=15.5cm, height=6.5cm]{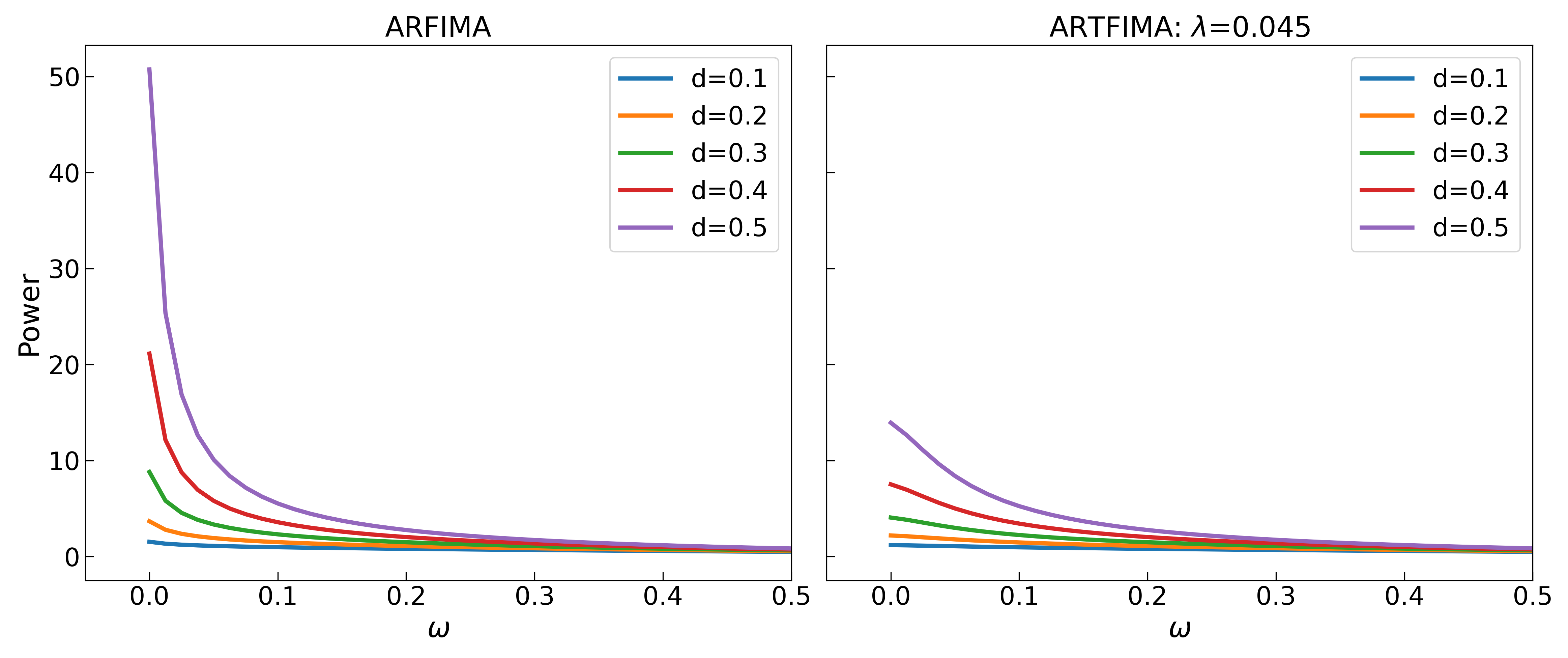}
    \caption{Spectral densities of ARFIMA and ARTFIMA models with $p=1$ and $q=0$. For both processes, $\phi=0.5$ and ARTFIMA $\lambda =0.045$.}  
    \label{fig: spectral}
\end{figure}

Let $\omega$ denote an angular frequency. The spectral density $f(\omega)$ is the Fourier transform of the covariance function and shows how the variance of the time series is distributed among the frequencies $-\pi \leq \omega  \leq \pi$. The spectral density when $\eta_t$ follows the ARTFIMA process in \eqref{eq:ARTFIMA_DLR} is \citep[Theorem 2.5 (a)]{sabzikar2019parameter} 
\begin{align}
\label{eq:spectraldens_ARTFIMA}
  f(\omega; \v \vartheta) & =\frac{\sigma_\varepsilon^2}{2\pi}\left |1-e^{-(\lambda+\mathrm{i}\omega)}\right |^{-2d}\left |\frac{\psi_q(e^{-\mathrm{i} \omega})}{\phi_p(e^{-\mathrm{i}\omega})}\right |^2,    
\end{align}
with parameter vector $\v \vartheta= (\phi_1, \dots, \phi_p, \psi_1, \dots, \psi_q, d, \lambda, \sigma^2_\varepsilon)^\top$, where $\sigma^2_\varepsilon=\mathrm{Var}(\varepsilon_t)$ and $\mathrm{i}^2 = -1$. The seasonal ARTFIMA($p,d,q)(P,D,Q)_s$, with $d \notin \mathbb{Z}$ and integer $D$,
\begin{align}
\phi_p(B)\phi_P^\star(B^s)\Delta^{d, \lambda} \Delta^D_s\eta_t = \psi_q(B)\psi_Q^\star(B^s)\varepsilon_t , \label{eq:ARTFIMA_DLR_seasonal}
\end{align}
where all quantities have been previously defined. The spectral density of $\Delta^D_s \eta_t$ in \eqref{eq:ARTFIMA_DLR_seasonal} is
\begin{align}
\label{eq:spectraldens_seasonal_ARTFIMA}
  f(\omega; \v \vartheta) & =\frac{\sigma_\varepsilon^2}{2\pi}\left |1-e^{-(\lambda+\mathrm{i}\omega)}\right |^{-2d}\left |\frac{\psi_q(e^{-\mathrm{i}\omega})\psi^\star_Q(e^{-\mathrm{i}s\omega})}{\phi_p(e^{-\mathrm{i}\omega})\phi^\star_P(e^{-\mathrm{i}s\omega})}\right |^2.    
\end{align}
Figure \ref{fig: spectral} shows the spectral density for both the ARFIMA and ARTFIMA process for different values of the fractional differencing parameter $d$. An important difference that we have stressed is the limiting behaviour as $\omega \rightarrow 0$, where the spectral density of the ARFIMA process diverges when $d>0$, whereas that of the ARTFIMA process does not. When estimating the power for real data, it is often observed to be bounded as discussed in Section \ref{section: DLR introduction}. Hence, the ARTFIMA model provides a better fit for small frequencies; see, for example, Figure \ref{fig: loglog gefcom}.

\section{Methodology}\label{section: DLR methodology}

\subsection{Frequency domain likelihood} \label{subsection: DLR frequency domain likelihood}

The frequency domain approach to inference relies on the asymptotic properties of the frequency representation of the time series process in \eqref{eq:DLR}. \cite{Engle1974band} shows how to rewrite \eqref{eq:DLR} in terms of the periodograms of $Y_t$ and $\v X_t$ and derive the ordinary least squares estimator of the transformed regression, which is also shown to be the best linear unbiased estimator. Our Bayesian approach requires a log-likelihood function, which can be derived using asymptotic properties of the periodogram data, and is outlined below. 

Let $Y_t$ and $\v X_t$ be zero-mean processes. Suppose that $\v \beta$ is known and define the pseudo data 
\begin{align}\label{eq:pseudo_data}
  Z_t = Y_t - \v X_t^\top \v \beta,\,\, t=1, \dots, T.  
\end{align}
The non-seasonal ARTFIMA process has the spectral density in \eqref{eq:spectraldens_ARTFIMA}, and when $\lambda = 0$ and $d=0$, it simplifies to the spectral density for a non-seasonal ARMA model. The general case assumes that $\eta_t$ is a seasonal ARTFIMA process and has the spectral density given by \eqref{eq:spectraldens_seasonal_ARTFIMA}. Similarly, the spectral density of the seasonal ARMA process is obtained by setting $\lambda = 0$ and $d=0$.

Recall $\omega \in [-\pi, \pi]$ as the angular frequency and denote the natural Fourier frequencies as $\omega_k = \frac{2\pi k}{T}$ for $k\in \mathcal{K}$, where 
\begin{equation} \label{eq: frequencies}
\mathcal{K}=
\begin{cases}
-\frac{T}{2}, -\frac{T}{2}+1, \dots, \frac{T}{2} - 1, \text{ if } T \text{ is even}, \\
-\frac{(T-1)}{2}, -\frac{(T-1)}{2}+1, \dots, \frac{(T-1)}{2}, \text{ if } T \text{ is odd}.
\end{cases}
\end{equation}
The frequency representation of the time series process $Z_t$ is obtained via its discrete Fourier transform (DFT), which is the complex-valued transform
\begin{align}
  J_Z(\omega_k) &  = \sum_{t=1}^T Z_t \exp(-\mathrm{i}\omega_kt),  \label{eq:DFT}
\end{align}
and can be computed for all $T$ frequencies $\omega_k$, $k \in \mathcal{K}$, using $\mathcal{O}(T\log_2(T))$ operations via the fast Fourier transform \citep{cooley1965algorithm}. Since the pseudo data depends on $\v \beta$, the DFT in \eqref{eq:DFT} needs to be recomputed for each new sample in the Markov chain Monte Carlo (MCMC) algorithm. However, following \cite{matsuda2009fourier}, 
\begin{align*}
  J_Z(\omega_k) & =   \sum_{t=1}^T (Y_t  - \v X_t^\top \v \beta)  \exp(-\mathrm{i}\omega_kt) \\
              & =   \sum_{t=1}^T Y_t \exp(-\mathrm{i}\omega_kt)  -   \left(\sum_{t=1}^T \v X_t \exp(-\mathrm{i}\omega_kt)\right)^\top  \v \beta  \\
              & = J_Y(\omega_k) - \v J_{\v X}(\omega_k)^\top \v \beta,
\end{align*}
where $J_Y$ is the DFT of $Y_t$, and $ \v J_{\v X}(\omega_k)$ is the $m$-dimensional column vector with the $j$th element being the DFT of the (univariate) time series $X_{jt}$. Thus, since $J_Y(\omega_k)$ and $\v J_{\v X}(\omega_k)$ only depend on the data $Y_t$ and $\v X_t$, and can be pre-computed before the MCMC algorithm, $J_Z(\omega_k)$ can be evaluated in $\mathcal{O}(T)$ for all frequencies when $\v \beta$ changes (assuming $m\ll T$).

Let $\Re(J_Z(\omega_k))$ and $ \Im(J_Z(\omega_k))$ be the real and imaginary parts of the DFT $J_Z(\omega_k)$ which is a weighted complex valued sum of the pseudo data. \citet[Theorem 2.1]{peligrad2010central} show that under some regularity conditions, 
\begin{align}
  \frac{1}{\sqrt{T}}(\Re(J_Z(\omega_k)), \Im(J_Z(\omega_k))), \quad T\rightarrow\infty,  \label{eq:asymptotic_dist_DFT}
\end{align}
converge in distribution to a bivariate normal distribution with (asymptotically) independent components having expected value 0 and variance $\pi f(\omega_k)$, with $f$ being the spectral density of $Z_t$. Moreover, they show that the  $\frac{1}{\sqrt{T}}J_Z(\omega_k)$ are asymptotically independent for all $k\in \mathcal{K}$.

Define the periodogram 
\begin{equation} \label{eq: DLR periodogram}
    I_Z(\omega_k) = \frac{1}{2\pi}\left|\frac{1}{\sqrt{T}}J_Z(\omega_k)\right|^2. 
\end{equation}
Then, for $k \in \mathcal{K}\setminus 0$, by \eqref{eq:asymptotic_dist_DFT},
\begin{align}
    \frac{I_Z(\omega_k)}{f(\omega_k)} & = \frac{1}{2}\left|\frac{1}{\sqrt{\pi f(\omega_k)}}\frac{1}{\sqrt{T}}J_Z(\omega_k)\right|^2 \sim \frac{\chi^2(2)}{2}, \label{eq:asymptotic_dist_scaled_periodogram}
\end{align}
as $T\rightarrow \infty$, where $\chi^2(\nu)$ denotes the chi-squared distribution with $\nu$ degrees of freedom. Note that $\nu = 2$ follows from $\left|\cdot\right|^2$ in \eqref{eq:asymptotic_dist_scaled_periodogram} being a sum of two squared (asymptotically) independent standard normal random variables when $k \in \mathcal{K}\setminus 0$. Since $\frac{\chi^2(2)}{2}$ is a standard exponential random variable, it follows that
\begin{align}
    I_Z(\omega_k) & \sim \mathrm{Exp}(f(\omega_k)),\,\, k \in \mathcal{K}\setminus 0,\label{eq:asymptotic_dist_periodogram}
\end{align}
where $\mathrm{Exp}(f(\omega_k))$ denotes an exponential distribution with mean $f(\omega_k)$. Emphasising the dependence on the parameters, the log-density of \eqref{eq:asymptotic_dist_periodogram} is
\begin{align}
    \log p(I_Z(\omega_k;\v \beta)|\v \theta) & = - \log(f(\omega_k; \v \vartheta)) - \frac{I_Z(\omega_k;\v \beta)}{f(\omega_k; \v \vartheta)}, \,\, k \in \mathcal{K} \setminus 0,\label{eq:log_dens_periodogram}
\end{align}
where $\v \theta = (\v \vartheta^\top, \v \beta^\top)^\top$ contains all unknown parameters. When $k=0$, $J_Z(0)=0$ since $J_Y(0)=\sum_{t=1}^T Y_t=0$ and  $\v J_{\v X}(0)=\sum_{t=1}^T  \v X_t= \v 0$ for demeaned data. 

The asymptotic distributions of the periodogram data underlie the idea of the so-called Whittle log-likelihood \citep{whittle1953estimation}: use the log-density of each periodogram observation $I_Z(\omega_k;\v \beta)$ in  \eqref{eq:log_dens_periodogram} to form the log-likelihood of all periodograms, which becomes a sum due to the (asymptotic) independence. For a real-valued process $Z_t$, both the periodogram and spectral density are symmetric around the origin; hence, only non-negative frequencies are considered. The Whittle log-likelihood is obtained by adding (due to the asymptotic independence) the log-densities in \eqref{eq:log_dens_periodogram} for all the positive frequencies (zeroth frequency excluded with demeaned data). The Whittle log-likelihood is, for odd $T$,
\begin{align}
  \ell_W(\v \theta) & =  -\sum_{k=1}^{(T-1)/2}  \left(\log(f(\omega_k; \v \vartheta)) + \frac{I_Z(\omega_k; \v \beta)}{f(\omega_k; \v \vartheta)} \right), \label{eq:Whittle_log_likelihood}
\end{align}
and the summation runs to $T/2-1$ instead if $T$ is even. \cite{guyon1982parameter} and \cite{kent1996spectral} show that the Whittle log-likelihood is a spectral approximation of the Gaussian time domain log-likelihood. In particular, they derive probabilistic convergence results of, respectively, the Whittle log-likelihood and its derivative, to the Gaussian log-likelihood and its derivative.

Finally, we stress that a frequency domain approach via the Whittle log-likelihood for dynamic linear regression models is inappropriate when using the ARFIMA model, because the Whittle approximation fails to hold for long memory processes, in particular for the small frequencies \citep{robinson1995log, Rousseau2012bayesnonparametric}. Thus, to carry out inference in dynamic linear regression models with ARFIMA errors, it is necessary to use the time domain log-likelihood, which is considerably slower to compute compared to a frequency domain approach that uses the ARTFIMA process; see Sections \ref{subsection: simulated data MCMC}, \ref{subsec:gefcom_application}, and \ref{sec: vic elec}. 

\subsection{Bayesian inference via Markov chain Monte Carlo}\label{subsec:Bayes_and_MCMC}
An important ain in time series is to learn the posterior distribution of $\v \theta = (\v \vartheta^\top, \v \beta^\top)^\top\subseteq \v \Theta \subset \mathbb{R}^{\dim_{\v \vartheta} + \dim_{\v \beta}}$ given realisations of the time series processes $Y_t\in \mathbb{R}$ and $\v X_t \in \mathbb{R}^{\dim_{\v X}}$.
Let $p(\v Z |\v \theta)$ denote the likelihood function of $\v \theta$ given the pseudo data $\v Z = (Z_1, \dots, Z_T)^\top$, which depends on the subset $\v \beta$ of the parameter vector $\v \theta$, but we suppress this dependence for notational simplicity. When the likelihood is obtained via the Whittle log-likelihood in \eqref{eq:Whittle_log_likelihood}, \ $p(\v Z |\v \theta)=\exp\left(\ell_W(\v \theta)\right)$. The likelihood function in this case is given the periodogram ordinates data $\v I = (I(\omega_0), \dots, I(\omega_{(T-1)/2}))^\top$, but since they are functions of $\v Z $ we keep using the notation $p(\v Z |\v \theta)$ (rather than $p(\v I|\v \theta)$).

The posterior distribution of $\v \theta$ is obtained using Bayes' theorem,
\begin{align}
  p(\v \theta |\v Z) & = \frac{p(\v Z|\v \theta)p(\v \theta)}{ p(\v Z)}, \,\, p(\v Z) = \int_{\v \Theta} p(\v Z|\v \theta)p(\v \theta)d \v \theta, \label{eq:Bayes_theorem}  
\end{align}
where $p(\v \theta)$ is the prior distribution of $\v\theta$. Markov chain Monte Carlo methods are a class of iterative procedures to sample from \eqref{eq:Bayes_theorem}, with the Metropolis-Hastings algorithm \citep{metropolis1953equation, hastings1970monte} arguably being the most popular. The Metropolis-Hastings algorithm constructs a Markov chain $\left\{ \v \theta^{(j)}\right\}$ by starting at some initial value $\v \theta^{(0)}$ and then, recursively, proposes a candidate draw $\v \theta^\prime$ from a proposal density $q(\v \theta |\v \theta^{(j-1)})$ and sets $\v \theta^{(j)}=\v \theta^\prime$ with acceptance probability 
\begin{align}
\alpha_{\mathrm{MH}} = \min\left(1, \frac{p(\v Z|\v \theta^\prime) p(\v \theta^\prime)/ q(\v \theta^\prime |\v \theta^{(j-1)})}{p(\v Z|\v \theta^{(j-1)})p(\v \theta^{(j-1)})/q(\v \theta^{(j-1)}|\v \theta^\prime )} \right).
    \label{eq:acc_prob}
\end{align}
If a proposed draw is rejected, $\v \theta^{(j)}=\v \theta^{(j-1)}$. The proposal density is often a random walk, e.g. 
\begin{align}\label{eq:proposal_dens}
q(\v \theta| \v \theta^{(j-1)}) & = \mathcal{N}(\v \theta|\v \theta^{(j-1)}, c \v \Sigma_{\mathrm{prop}}),
\end{align}
where $\mathcal{N}(\v x|\v \mu, \v \Sigma)$ denotes the multivariate normal density of $\v x$ with mean $\v \mu$ and covariance matrix $\v \Sigma$. Moreover,
 $\v\Sigma_{\mathrm{prop}}$ is an approximation of the posterior covariance matrix, for example minus the inverse of the observed Fisher information evaluated at the maximum a posteriori (MAP) estimate, and $c = 2.38/\sqrt{\dim(\v \theta)}$ for optimality, resulting in $\alpha_{\mathrm{MH}} = 0.234$ \citep{gelman1997weak}. However, in the presence of intricate geometry in the posterior due to complex dependence structures, selecting a proposal covariance as described may yield low acceptance probabilities. An alternative approach that we implement is to adaptively tune the proposal covariance matrix \citep{haario2001adaptive}.  This technique uses the information accumulated thus far through the previous draws to adjust the covariance proposal matrix. The adaptive approach can lead to more accurate simulated approximations of the target posterior and less auto-correlation in the resulting chain. To target a specified overall acceptance rate of the sampler (0.234), we follow \cite{garthwaite2016adaptive} who utilise a Robbins-Monro search process to compute the scaling constant $c$ in \eqref{eq:proposal_dens} of the adapted proposal covariance matrix.

The acceptance probability in \eqref{eq:acc_prob} is set so that, informally speaking, the Markov chain consists of samples from the posterior in \eqref{eq:Bayes_theorem} after a warm-up period referred to as the burn-in. Let $\left\{ \v \theta^{(j)}\right\}_{j=1, \dots, N}$ be the samples after discarding the burn-in. If the algorithm rejects too often, the Markov chain (the samples) becomes ``sticky'', which causes inefficient estimates of posterior expectations. To explain how this inefficiency is measured, suppose $\theta$ is scalar-valued and consider estimating the expectation of a function $h$, i.e.\
\begin{align}
  \mathrm{E}(h(\theta)) = \int_{\v \Theta} h(\theta) p(\theta |\v Z) d\theta. \label{eq:posterior_expectation}  
\end{align}
By the weak law of large numbers,
\begin{align}
  \widehat{\mathcal{I}}_N & = \frac{1}{N} \sum_{i=1}^N h(\theta^{(j)}) \overset{P} \rightarrow \mathrm{E}(h(\theta)). \label{eq:law_of_large_numbers}
\end{align}
If the samples are independent, then the asymptotic variance of $\sqrt{N} \widehat{\mathcal{I}}_N$ is $\sigma_h^2$, where $\sigma^2_h = \mathrm{Var}(h(\theta))$. However, MCMC results in correlated samples and then the asymptotic variance of $\sqrt{N} \widehat{\mathcal{I}}_N$ is $\sigma_h^2(1 + \sum_{i=1}^\infty \rho_i)$, where $\rho_i$ is the autocorrelation at the $i$th lag the Markov chain. The term $(1 + \sum_{i=1}^\infty \rho_i)$ is called the inefficiency factor. It measures how much the asymptotic variance of the estimate of a posterior expectation is inflated compared to an ideal sampler that produces independent draws. The effective sample size (ESS) of the resulting Markov chain is defined as $\text{ESS} = N / (1 + \sum_{i=1}^\infty \rho_i)$ (note that $\text{ESS}=N$ with independent samples). Thus, the effective sample size estimates the number of samples equivalent to the number of independent samples the sampler is generating, see 
 \cite{geyer2011introduction} for more details.

\subsection{Parameterisations and prior distributions}
The autoregressive parameters $\phi_i$ are reparameterised in terms of partial autocorrelations $-1 < \widetilde{\phi}_i < 1 $, for $i=1,\dots,p$, following \cite{barndorff1973parametrization} to ensure the process is always stationary. For the moving average parameters $\psi_j$, for $j=1,\dots,q$, the same reparameterisation in terms of $\widetilde{\psi}_j$ ensures the process is always invertible and the moving average parametrisation is unique. To make the variance parameter $\sigma^2_\varepsilon$ and the tempering parameter $\lambda$ unrestricted, we reparameterise using log-transformations. For the ARTFIMA process, the fractional parameter $d$ ($\notin \mathbb{Z}$) is unrestricted (since stationarity does not depend on $d$). For ARFIMA models, on the other hand, $-0.5<d<0.5$ is a necessary condition for stationarity. We thus reparameterise $d$ to an unrestricted parameter $\widetilde{d}$ by the scaled Fisher transformation $\widetilde{d}=\mathrm{arctanh}(2d)$.

When using a sample from the multivariate normal proposal as described in Section \ref{subsec:Bayes_and_MCMC}, a prior on the interval $(-1,1)$ for each autoregressive and moving average parameter ensures $|\widetilde{\phi}_i| < 1$ and $|\widetilde{\psi}_j| < 1$ for the $i$ and $j$ above. We set independent (within and between) uniform priors $\v \phi \sim \mathrm{Unif}(-1,1)^p$ and $\v \psi \sim \mathrm{Unif}(-1,1)^q$. The priors for $\log(\sigma^2_\varepsilon)$ and $\log(\lambda)$ are $\mathcal{N}(0,100)$. For the ARTFIMA process $d\sim \mathcal{N}(0,1)$, while for ARFIMA models, the prior $\widetilde{d}\sim \mathcal{N}(0,1)$ implies a weakly informative on $(-0.5, 0.5)$ for $d$. Finally, $\beta_k$, for $k=1,\dots, m$, are assigned independent $\mathcal{N}(0,100)$ priors. We have verified that our results are not influenced by the choice of prior (because the datasets studied are very large).

\section{Simulation studies} \label{section: DLR simulated example}

\subsection{Accuracy of the approximation of the time domain likelihood} \label{subsection: simulated data MCMC}
We compare posterior inference via the Whittle likelihood to those using the exact Gaussian and Kalman filter likelihoods under the same weakly informative priors for dynamic linear regression models on simulated data. By the exact Gaussian likelihood, we mean evaluating a multivariate normal density where the covariance matrix is computed from the autocovariance function. The Kalman filter likelihood performs the analytic filtering steps recursively to compute the likelihood \citep{hamilton2020time}. Unlike the Gaussian likelihood approach, the Kalman filter avoids inverting a large covariance matrix. Note that both methods use the same (exact) time-domain likelihood, but are evaluated differently. This example is restricted to an ARMA process as the ARFIMA/ARTFIMA likelihoods cannot be efficiently evaluated using the Kalman filter (recall from Section \ref{section: DLR introduction} that they do not have finite-dimensional state space representations).

Let $y_t = \beta_1x_t + \eta_t$ where $\eta_t$ is an ARMA(3,1) model with $T=5{,}001$. We simulate using the true value $\beta_1 = \beta^{(0)}_1 = 3$ and the reparameterised vector ($\psi_1^{(0)} = \widetilde{\psi}_1^{(0)}$) $$\v \vartheta^{(0)} = \left(\widetilde{\phi}_1^{(0)}, \widetilde{\phi}_2^{(0)}, \widetilde{\phi}_3^{(0)}, \psi_1^{(0)}, \log(\sigma_{\varepsilon}^{2(0)})\right)^\top = \left(0.4, -0.2, 0.1, 0.2, \log(2)\right)^\top, $$
and with $\left(\phi_1^{(0)}, \phi_2^{(0)}, \phi_3^{(0)}\right)^\top = \left(0.5, -0.248, 0.1\right)^\top$ in the original parameterisation, where the superscript $^{(0)}$ denotes a true value.
The exogenous predictor $x_t$ is assumed to be known and is simulated from an ARMA(2,2) process. The MCMC algorithm described in Section \ref{subsec:Bayes_and_MCMC} is used to sample the three posteriors for $10{,}000$ iterations with a burn-in of $3{,}000$ samples. Table \ref{tbl: DLR whittle vs gaussian repeated simulation} reports the mean squred error (MSE) of the posterior mean for $1{,}000$ replicates of the data for each parameter. The MSE of both the Whittle likelihood and the Gaussian likelihood are small and the differences are likely attributed to Monte Carlo error.

\begin{table}[ht]
\centering
\begin{tabular}{c c c c c c c}
\hline 
\textbf{MSE}  &  $\beta_1$ & $\phi_1$ & $\phi_2$ & $\phi_3$ & $\theta_1$ & $\sigma^2$ \\
\hline
Gaussian & 0.013 & 0.076  & 0.007 & 0.003 & 0.072 & 0.002 \\
\hline
Whittle  & 0.014 & 0.065 & 0.006 & 0.002 & 0.063 & 0.002  \\
\hline
\end{tabular}
\caption{The mean square error (MSE) for posterior means (reparameterised) with the Gaussian likelihood and Whittle likelihood under the same prior for $1{,}000$ data simulation replicates.}
\label{tbl: DLR whittle vs gaussian repeated simulation}
\end{table}

\begin{figure}[htp]
\includegraphics[width=14.5cm, height=7cm]{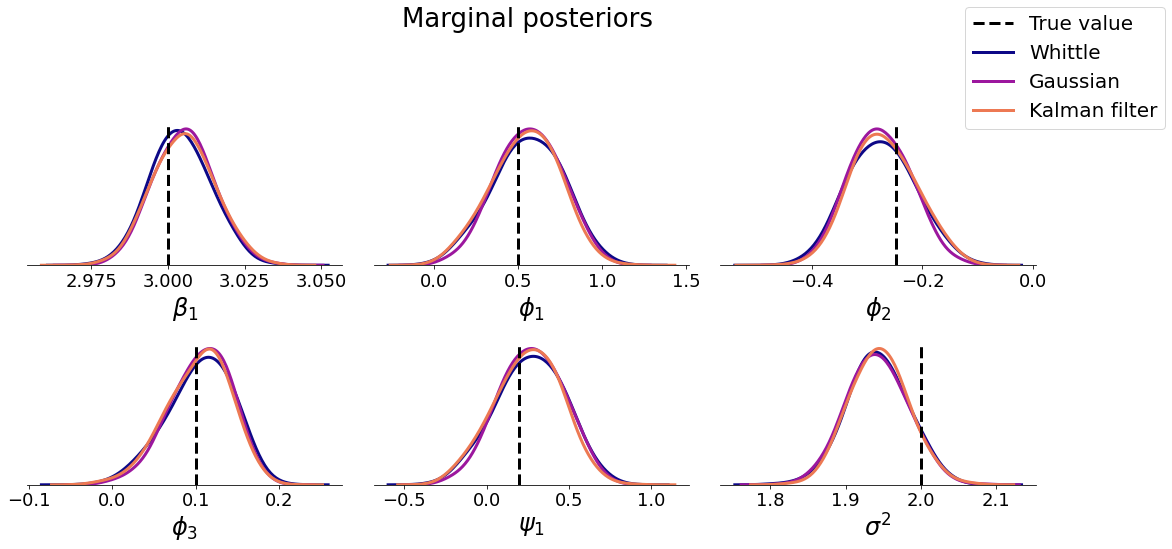}
\centering
\caption{Kernel density estimates of the marginal Whittle, Kalman filter and Gaussian posteriors (original parameterisation) for a DLR model with $\eta_t \sim$ ARMA(3,1).}
\label{fig: DLR true vs whittle marginal posterior}
\end{figure}

Figure \ref{fig: DLR true vs whittle marginal posterior} plots kernel density estimates of the marginal posteriors for each parameter given data from one simulation. As shown, the three posteriors are almost identical. Hence, the $n=2{,}500$ periodogram ordinates were enough to get a very close approximation of the Whittle likelihood to the true time domain likelihood.

\begin{figure}[ht]
\includegraphics[width=14.5cm, height=7cm, keepaspectratio]{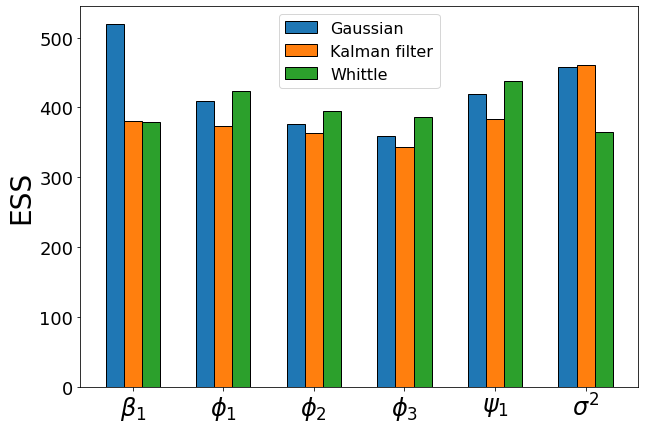}
\centering
\caption{Effective sample sizes for $10{,}000$ MCMC iterations for each parameter (original parameterisation) in a DLR model with $\eta_t\sim \mathrm{ARMA}(3,1)$. The results are shown for the 3 methods in the legend.}
\label{fig: DLR ess simulate data}
\end{figure}

Figure \ref{fig: DLR ess simulate data} reports the effective sample sizes. The figure shows that the Gaussian likelihood has a $\approx 25\%$ higher ESS for the $\beta_1$ parameter compared to the Whittle likelihood and the Kalman filter likelihood. The Whittle approach has a lower ESS for $\sigma^2$ compared to its time domain counterparts. The AR and MA parameters are comparable for all three posteriors, with the Whittle posterior having slightly higher ESS for all.

Table~\ref{tbl: DLR simulated data run times} presents the run times (seconds) of each method. The proposed Whittle approach is roughly 8 times faster than the Gaussian likelihood and 94 times faster than the Kalman filter.
The Gaussian likelihood uses the \texttt{SuperGauss} package in R \citep{ling1supergauss}, which scales as $\mathcal{O}(T \log_2^2 (T))$ using the superfast Toeplitz algorithm proposed in \cite{ammar1988superfast}. Despite the Kalman filtering being performed in $\mathcal{O}(T)$ operations, the algorithm sweeps through all observations recursively for each new MCMC proposal and is thus time-consuming, especially in the Python implementation used in this paper. The Whittle log-likelihood \eqref{eq:Whittle_log_likelihood}, on the other hand, is a simple sum that only requires computing $I_Z(\omega)$ in \eqref{eq: DLR periodogram} and the relevant spectral density, and is thus much faster than the Kalman filter.

\begin{table}[ht]
\centering
\begin{tabular}{c c c c}
\hline
\textbf{Run time}  & Whittle & Gaussian & Kalman filter \\
\hline
total (s)          & 10.0  & 80.54 & 942.58 \\
\hline
\end{tabular}
\caption{Computation time (seconds) for sampling $10{,}000$ MCMC draws for each method.}
\label{tbl: DLR simulated data run times}
\end{table}

\begin{figure}[t]
\includegraphics[width=15.5cm, height=20cm, keepaspectratio]{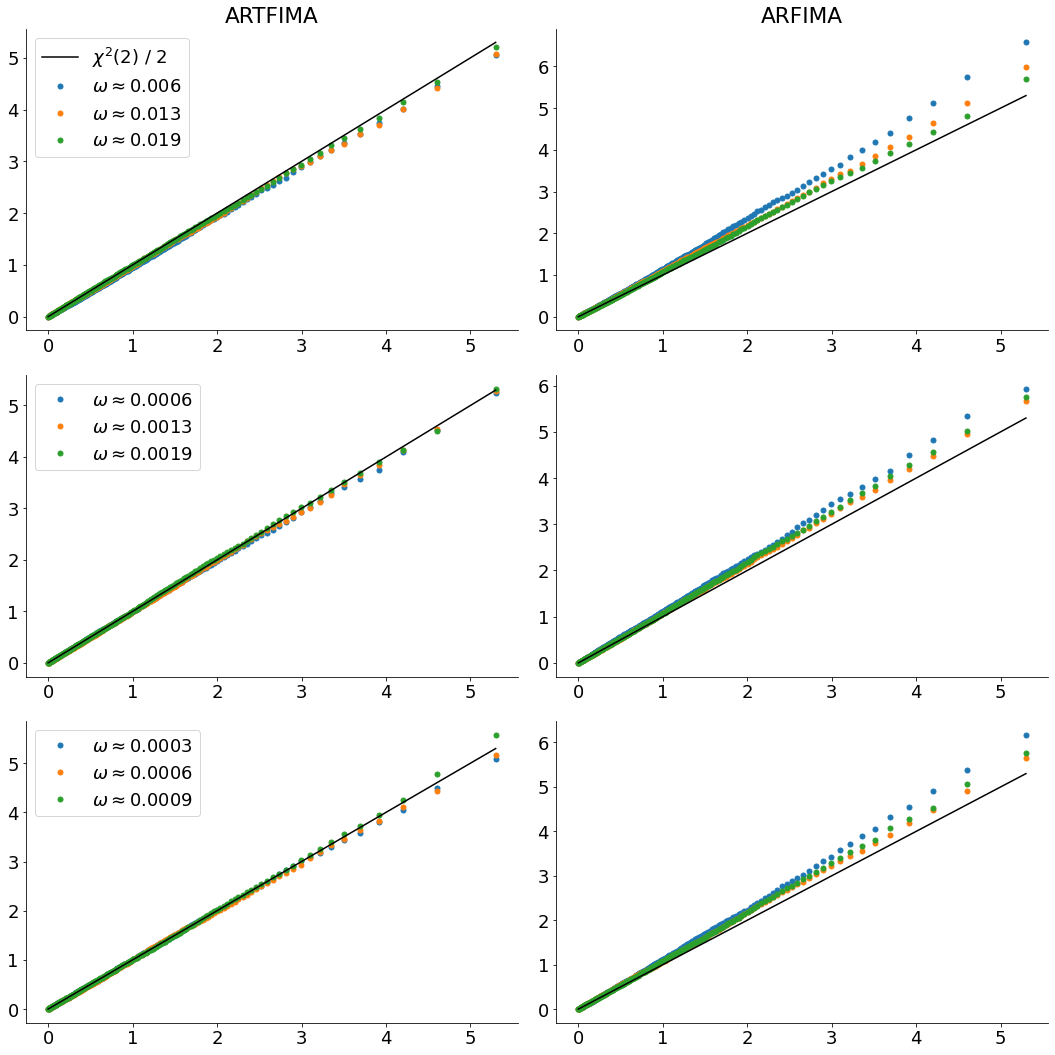}
\centering
\caption{Quantile-quantile (QQ) plots of the ratio ($\widehat{\beta}_1$ varying over replicates) $I_Z(\omega_k;\widehat{\beta}_1) / f(\omega_k;\v\vartheta^{(0)})$  for simulated DLR ARTFIMA (left panel) vs DLR ARFIMA (right panel) models for the three lowest positive frequencies. The top row, middle row, and bottom row, correspond to $T=1{,}001$, $T=10{,}001$, and $T=20{,}001$, respectively.}
\label{fig: arfima vs artfima qq plots}
\end{figure}

\subsection{The periodogram distribution for low frequencies} \label{sec: DLR periodogram simulations}

This section empirically verifies the periodogram distribution for small frequencies for the DLR with ARTFIMA errors. We also illustrate how this fails in the case of an ARFIMA error process. In a non-DLR setting, i.e.\ $\v \beta = \mathbf{0}$, the latter has been investigated theoretically in \cite{robinson1995log, Rousseau2012bayesnonparametric} and empirically in \cite{meerschaert2014tempered, sabzikar2019parameter}. The aim of this simulation study is to empirically verify that the ARTFIMA scaled low-frequency periodogram observations of the pseudo data in \eqref{eq: DLR periodogram} follows \eqref{eq:asymptotic_dist_scaled_periodogram}, and that it is violated in the ARFIMA case.

Consider the models $y_t = \beta_1 x_t + \eta^{(i)}_t$ for $i=1,2$, $$\eta^{(1)} \sim \mathrm{ARTFIMA}(2, d, \lambda, 0),\,\, \eta^{(2)} \sim \mathrm{ARFIMA}(2, d, \lambda, 0),$$
and  $x_t \sim \text{ARMA(1,1)}$ (fixed throughout the simulation and the same for both processes). We simulate the models with $\beta_1 = \beta_1^{(0)} = 0.1$ and process parameters:
\begin{enumerate}
    \item[$(i=1)$] $\v \vartheta^{(0)}=(\phi_1^{(0)}, \phi_2^{(0)}, d^{(0)}, \lambda^{(0)}, \sigma^{2(0)})^\top = (0.742, 0.227, 2.139, 0.616, 1)^\top $,
    \item[$(i=2)$] $\v \vartheta^{(0)}=(\phi_1^{(0)}, \phi_2^{(0)}, d^{(0)}, \sigma^{2(0)})^\top = (1.466, -0.525, 0.493, 1)^\top$.
\end{enumerate}
The parameter values of the error processes are chosen to match those obtained by the real data application in Section \ref{sec: vic elec}. Recall from \eqref{eq:asymptotic_dist_scaled_periodogram} that the ratio of the periodogram and its spectrum follows the distribution
\begin{equation} \label{eq: ratio I / f}
    \frac{I_Z(\omega_k; \beta_1)}{f(\omega_k;\v \vartheta)} \sim \frac{\chi^2(2)}{2},
\end{equation}
as $T \rightarrow \infty$ for $\omega_k = \frac{2\pi k}{T}, k \in \mathcal{K}$, where  $\mathcal{K}$ is defined in \eqref{eq: frequencies} and the dependence of the parameters is emphasised. Note that if we set $\beta_1=\beta_1^{(0)}$ when computing the periodogram, then this corresponds to the known results in a non-DLR setting since the pseudo data in \eqref{eq:pseudo_data} by the construction of the simulation follows an ARTFIMA (or ARFIMA) process. To obtain $\beta_1 \neq \beta_1^{(0)}$, we evaluate the periodogram at the MAP $\widehat{\beta}_1$ in each replication. We verify \eqref{eq: ratio I / f} with $\v \vartheta = \v \vartheta^{(0)}$ and $\beta_1 = \widehat{\beta}_1$ by simulating $10{,}000$ realisations for $T = 1{,}001, 10{,}001, 20{,}001$ and keeping the three smallest positive Fourier frequencies $\omega_k$ (excluding zero). Figure \ref{fig: arfima vs artfima qq plots} displays the quantile-quantile (QQ) plots of $I_Z(\omega_k; \widehat{\beta}_1) / f(\omega_k; \v \vartheta^{(0)})$ against $\chi^2_2/2$ for all three cases of $T$. The left and right panels display the plots for the ARTFIMA and ARFIMA models, respectively. The upper, middle, and bottom rows correspond to $T = 1{,}001$, $T=10{,}001$ and $T=20{,}001$, respectively. The left panel (ARTFIMA) of Figure \ref{fig: arfima vs artfima qq plots} shows that for the three lowest frequencies, the empirical quantiles of the periodogram ratios are almost identical to their theoretical counterpart for all $T$. This confirms the suitability of \eqref{eq: ratio I / f} for the small frequencies in the ARTFIMA model case. In contrast, the right panel (ARFIMA) shows that \eqref{eq: ratio I / f} is unsuitable for the small frequencies in the ARFIMA model case, and that an increasing $T$ does not make \eqref{eq: ratio I / f} more suitable.

The violation of \eqref{eq: ratio I / f} is partly attributed to the fact that the spectrum of the ARFIMA model behaves as a power law and diverges as $\omega \rightarrow 0$. \cite{sabzikar2019parameter} and \cite{meerschaert2014tempered} show that for applications such as hydrology, finance, and climatology, the ARFIMA spectral density provides a poor fit of the periodogram at the low frequencies. We show the same issue in a dynamic linear regression model with ARFIMA errors; see Figure \ref{fig: loglog gefcom} in Section \ref{subsec:gefcom_application}. A pragmatic solution is to use a low-frequency cutoff in the estimation of ARFIMA models in the frequency domain; however, this incurs a substantial loss of information in the estimation of the long-memory parameter $d$. Section \ref{section: DLR applications} thus uses the exact Gaussian likelihood for dynamic linear models with ARFIMA errors, which corresponds to the exact maximum likelihood approach in \cite{doornik2004inference}. The resulting computation is much slower than a frequency domain approach and stresses the importance of our contribution, namely: a computationally fast dynamic linear model that accounts for long-range dependence in the error process.

\section{Applications} \label{section: DLR applications}

\subsection{Overview and objectives}
This section analyses electricity demand as the response variable conditional on relevant exogenous variables for two real-world datasets. The two datasets contain data for New England, United States of America (Example 1), and Victoria, Australia (Example 2), presented in Sections \ref{subsec:gefcom_application} and \ref{sec: vic elec}, respectively. Forecasting for Example 2 is considerably more challenging than for Example 1 and we show that accounting for long-range dependence is essential to obtain good forecast accuracy for moderately long horizons. In contrast, long-range dependency does not improve the forecasts in Example 1; however, we show it has a significant impact on the posterior distribution of the regression parameters $\v \beta$. Example 1 also illustrates the failure of a dynamic linear model with an ARFIMA error process to fit the observed power spectrum at low frequencies. 

We compare the estimation times and forecasting performances of models with three types of error processes: ARTFIMA, ARFIMA, and ARMA. In addition, we include a neural-network-based time series forecasting model, neural hierarchical interpolation for time series forecasting (NHITS) \citep{NHITS_neural_model}, as an external benchmark. NHITS lies outside the probabilistic modelling framework considered in this paper and is therefore evaluated solely in terms of forecasting performance; as a consequence, the log-predictive density score is not reported for this model. The model is implemented using the \texttt{neuralforecast} Python package \citep{NeuralForecast} and uses the same set of exogenous variables as the dynamic linear regression models, which are assumed to be known at the time of forecasting.

Section \ref{subsec:model_settings} discusses the choice of model settings and Section \ref{subsec:forecast_eval_settings} presents the metrics used for the forecasting evaluation.

\subsection{Model settings}\label{subsec:model_settings}
We consider three types of error processes: ARTFIMA, ARFIMA and ARMA. To choose the order of $p$ and $q$, a search over the model space is considered up to a maximum of $p = 2, q = 1$. For various orders of $p$ and $q$, we select the model with the smallest deviance information criterion (DIC) value \citep{spiegelhalter2002bayesian}. For a given model $\mathcal{M}$, the deviance is defined as
\begin{equation}\label{eq:DIC}
    D(\bb{\theta}) = -2\log  p(\bb{y} | \bb{\theta}, \mathcal{M}) + 2\log h(\bb{y}),
\end{equation}
where $h(\bb{y})$ is a fully specified standardising term which is a function of only the data, which we set to be $h(\bb{y})=1$ for all models. The DIC consists of two components, 
\begin{equation}\label{eq:DIC2}
    DIC = \overline{D}(\bb{\theta}) + p_D,
\end{equation}
where the first term is the posterior expectation of the deviance in \eqref{eq:DIC},
\begin{equation*}
    \overline{D(\bb{\theta})} = \text{E}\left[ D(\bb{\theta})\right] = \text{E}\left[ -2\log p(\bb{y} | \bb{\theta}, \mathcal{M})\right].
\end{equation*}
The first term in \eqref{eq:DIC2} assesses how well the model fits the data, with smaller values being preferred. The second component, $p_D$, is the effective number of parameters of the model, which penalises the complexity of the model,
\begin{equation*}
    p_D = \overline{D(\bb{\theta})} - D({\v \theta}^\star) = \text{E}\left[ -2\log p(\bb{y} | \bb{\theta}, \mathcal{M})\right] + 2\log p(\bb{y} | {\v \theta}^\star, \mathcal{M}),
\end{equation*}
where ${\v \theta}^\star$ is the posterior mean. Thus, the DIC is a trade-off between the complexity and the adequacy of a model \citep{chan2016fast}.

Once the best $p$ and $q$ are chosen for the ARMA error process, we compare the ARMA model to the ARTFIMA and ARFIMA models with the same $p$ and $q$. Likewise, we find the best $p$ and $q$ for the ARTFIMA error process and compare it with the ARMA and ARFIMA models with the same $p$ and $q$. However, this is not done for ARFIMA since performing MCMC over the whole model space under the Gaussian likelihood is impractical with the large datasets in the applications.

It is well known (e.g.\ \citealp{wheeler2024likelihoodbasedinferencearma}) that the log-likelihood function for ARMA models may exhibit multimodality, for example when root or near root cancellation is present, and this is also true for ARTFIMA and ARFIMA. Moreover, the parameters $d$ and $\lambda$ introduce additional complexity to the geometry of the likelihood function. To avoid finding a poor local mode, we use global optimisation via Basin-hopping \citep{li1987monte} to find the MAP estimate, which serves as an initial starting value for MCMC. 

\subsection{Settings for forecast evaluation}\label{subsec:forecast_eval_settings}

For a given $p$ and $q$ determined as described in Section \ref{subsec:model_settings}, we perform time series leave-future-out-cross-validation to evaluate the forecasting performance. We follow \citet[Ch. 7, Ch. 10]{hyndman2018forecasting} and assume all exogenous variables $\mathbf{X}_t$ are known when forecasting. We use a sliding/rolling window approach with a fixed training size of $T$ with $k=100$ out-of-sample observations for testing. To compare models, we use three different metrics: log-predictive density score (LPDS), root mean square error (RMSE) and continuous rank probability score (CRPS). We compute each metric for different forecast horizons, $h = 1,\dots,15$ and note that the $h$-step ahead case results in $k-h+1$ out-of-sample observations ($k=100$). Each time the window is rolled forward (to predict the next out-of-sample observation), we re-estimate the model and forecast $h$-steps-ahead. However, for ARFIMA, the model is only estimated once on the initial training set; this is because re-estimating the model is very costly due to the evaluation of the Gaussian likelihood. This does not have a large impact since the sliding window approach only impacts the posterior negligibly due to the large amount of data. We note that approximate leave-future-out cross-validation techniques have been proposed \citep{burkner2020approximate} for computing the LPDS that do not require reestimating the posterior. The other metrics used in this paper, however, require the posterior samples and the cost of computing the LPDS is negligible given these samples. Hence we do not implement \cite{burkner2020approximate}.

The $h$-step-ahead log-predictive density score (LPDS) is
\begin{equation}
    \label{eq: LPDS conditional density}
    \text{LPDS}^{(h)} = \frac{1}{k-h+1} \sum^{k-h}_{i=0}\log p(y_{T+h+i} | y_{1+i:T+i}),
\end{equation}
where the $y_{T+h+i}$ for $i=0,\dots, k-h$ are the out-of-sample observations for which we evaluate the log-predictive density, 
\begin{align*} \label{eq: DLR: post pred time series}
\log p(y_{T+h+i}|y_{1+i:T+i}) &= \log\int_{\Theta} p(y_{T+h+i}|\btheta, y_{1+i:T+i}) p(\btheta | y_{1+i:T+i}) d\btheta, \\
&\approx \log \left( \frac{1}{M} \sum^M_{m=1} p(y_{T+h+i}|\btheta^{(m)}, y_{1+i:T+i}) \right),
\end{align*}
and the last line is a Monte Carlo approximation thereof, with $\bb{\theta}^{(m)}\sim p(\btheta | y_{1+i:T+i})$ and $M=900$. The LPDS gives a measure of how well the model fits the out-of-sample data \citep{gelman2014understanding}, with larger values being preferred. The other two measures (RMSE and CRPS), however, have a reciprocal scale (smaller is better), and thus, we report the negative LPDS so that smaller values are preferred for all three metrics.

To assess point forecasts from a given model, define the conditional expectation when predicting the $i$th out-of-sample observation as
\begin{equation*}
  \widehat{y}_{T+h+i} = \text{E} [ \widetilde{y}_{T+h+i} | y_{1+i:T+i}] = \int \widetilde{y}_{T+h+i} p( \widetilde{y}_{T+h+i} | y_{1+i:T+i}) d\widetilde{y}_{T+h+i}, \,\, i=0,\dots, k-h, 
\end{equation*}
where $p(\widetilde{y}_{T+h+i} | y_{1+i:T+i})$ is the posterior predictive distribution. To evaluate the performance of the point forecasts, the root mean square error (RMSE) is computed as 
\begin{equation*}
    \text{RMSE}^{(h)} = \sqrt{\frac{1}{k-h+1}\sum^{k - h}_{i=0} \left( y_{T+h+i} - \widehat{y}_{T+h+i}\right)^2},
\end{equation*}
for each forecast horizon $h$. The RMSE is the square root of the average squared distance between the out-of-sample observations and their forecasted values, and thus, smaller values are preferred.

The distributional forecasts $F()$ are assessed via the continuous rank probability score (CRPS), which is defined as
\begin{equation*}
    \text{CRPS}^{(h)}_i(F, y_{T+h+i}) = \int_{\mathbb{R}}\Big( F(\widetilde{y}_{T+h+i}) - \mathbf{1}\{ y_{T+h+i} \leq \widetilde{y}_{T+h+i}\} \Big) ^2 d\widetilde{y}_{T+h+i},
\end{equation*}
where $\mathbf{1}$ is the indicator function, taking value one if $ y_{T+h+i} \leq \widetilde{y}_{T+h+i}$ or zero otherwise \citep{matheson1976scoring}. When the distributional forecasts $F$ and the data distribution are equal, the CRPS achieves its minimum and thus smaller values are preferred. The distributional forecasts are unknown but can be estimated using samples from the $h$-step-ahead predictive posterior distribution to estimate the empirical cumulative distribution function $\widehat{F}(\widehat{y}_{T+h+i})$ for all, $i =0, \dots, k-h$. The final CRPS is formed by taking the mean over the out-of-sample observations, i.e.\
\begin{equation*}
    \text{CRPS}^{(h)} = \frac{1}{k-h+1}\sum^{k-h}_{i=0}\text{CRPS}^{(h)}_i(F, y_{T+h+i}).
\end{equation*}
To compute the RMSE and CRPS, we use 900 samples from the posterior predictive distribution.

\subsection{Example 1: New England electricity demand}\label{subsec:gefcom_application}

\begin{figure}[ht]
\includegraphics[width=15.5cm, height=10cm]{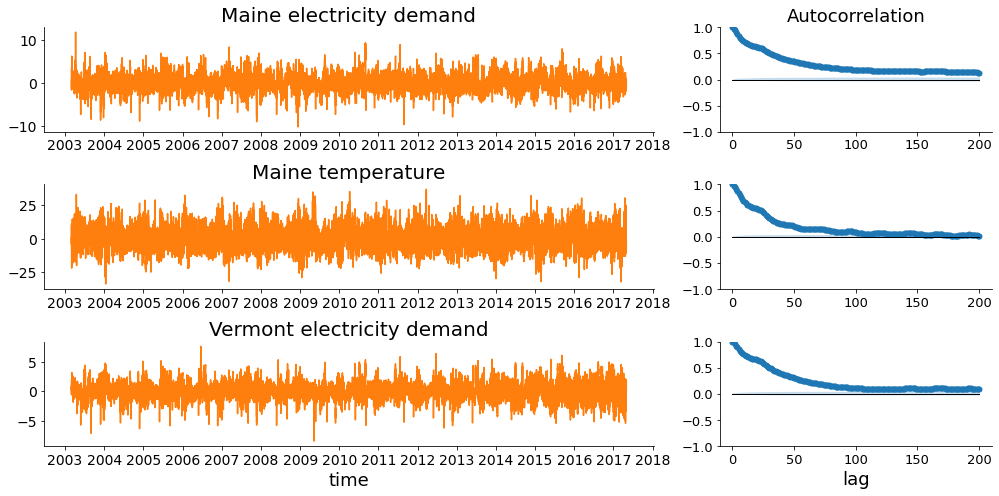}
\centering
\caption{The left panels plot the Maine and Vermont hourly electricity demand and temperature data each after removing multi-seasonal and trend components and applying a Box-Cox transformation. The right panels are the corresponding autocorrelation plots.}
\label{fig: gefcom data}
\end{figure}

The New England data consists of hourly electricity demand (megawatts) and temperature (degrees Celsius) for the states of Maine and Vermont from 1st March 2003 to April 30th, 2017, with $T=124{,}200$ observations presented in \cite{hong2016probabilistic}. The response, Maine electricity demand $y_t$, is modelled as a linear combination of lagged exogenous variables, Maine temperature $x_{1,t-1}$ and Vermont (neighbouring state) electricity demand squared $x^2_{2,t-1}$. Strong multi-seasonal and trend components exist within each variable. The \texttt{mstl} function in the R package \texttt{Forecast} \citep{hyndman2008automatic} is used to remove these effects. Figure \ref{fig: gefcom data} plots each variable after removing the multi-seasonal and trend components and applying a Box-Cox transformation \citep{box1964analysis}. The figure shows that the response exhibits some long memory, making it a suitable candidate for an ARFIMA or ARTFIMA error process.

The lowest DIC for ARMA and ARTFIMA error processes occurred for $p=2$ and $q=1$ in both cases. Table \ref{tbl: gefcom 2,1 models DIC and timings} reports the DIC values for the three models as well as the average time for one evaluation of the log-likelihood function. The table shows that ARTFIMA has the lowest DIC value, followed by ARFIMA, with ARMA having the highest (recall that lower is better). However, ARFIMA is the most computationally demanding model and is roughly 40 times slower than the other two models since the Whittle likelihood cannot be used. ARTFIMA and ARMA have comparable computational time per iteration of the log-likelihood of 11.5ms and 10.5ms, respectively.

\begin{table}[ht]
\centering
\begin{tabular}{l*{6}{c}r}
\hline 
        & ARTFIMA$(2, d, \lambda, 1)$ & ARFIMA$(2, d, 1)$ & ARMA$(2,1)$ \\
         
\hline
DIC        & $\bb{71987.56}$ & 72155.15 & 73030.91  \\
\hline
Time (ms) & 11.3 & 403.0 & 10.5 \\
\hline
\end{tabular}
\caption{DIC values and the average time (milliseconds) for one log-likelihood evaluation for each dynamic regression model for New England electricity demand. The lowest DIC value is in boldface.}
\label{tbl: gefcom 2,1 models DIC and timings}
\end{table}

Figure \ref{fig: forecast metrics gefcom 2,1} plots the negative LPDS, RMSE and CRPS for each model over the forecast horizon $h$. For the negative LPDS, ARFIMA performs the best here for $h > 3$, and the results for ARMA and ARTFIMA are almost identical. Hence, there is no visual distinction between the two. For the RMSE, NHITS achieves the lowest score at the one-step-ahead horizon ($h=1$), whereas the dynamic linear regression models perform better for longer forecast horizons ($h>1$). The ARFIMA model has the lowest RMSE score for $h>1$; however, when $h>6$, the ARMA and ARTFIMA models have lower RMSE scores than ARFIMA. The CRPS is similar for all dynamic linear regression models, with no such model clearly outperforming the others for several forecast horizons. NHITS attains the lowest CRPS for the one-step-ahead forecast ($h=1$), but is outperformed by the dynamic linear regression models at longer forecast horizons. A possible explanation is that neural-network-based forecasts may underestimate predictive, particularly when variability in the training data is limited \citep{hyndman2018forecasting}. Figure \ref{fig: h_1_6 gefcom 2,1} plots the predictions and prediction intervals for the three error processes when predicting the electricity demand one and six hours ahead, i.e.\ $h = 1, 6$. All three error processes perform accurate predictions, with the prediction intervals getting wider for larger horizons as expected. To summarise the results, we conclude that ARMA and ARTFIMA overall provide better scores and are computationally much more efficient than ARFIMA. ARMA performs almost identically to ARTFIMA, which indicates that the (semi) long-range dependency feature of ARTFIMA does not strongly influence the forecasts in this case. However, inspecting the posterior density of the regression parameter $\v \beta$ in Figure \ref{fig: beta_posterior 2,1} reveals a significant difference.  

\begin{figure}[ht]
\includegraphics[width=15cm, height=8cm, keepaspectratio]{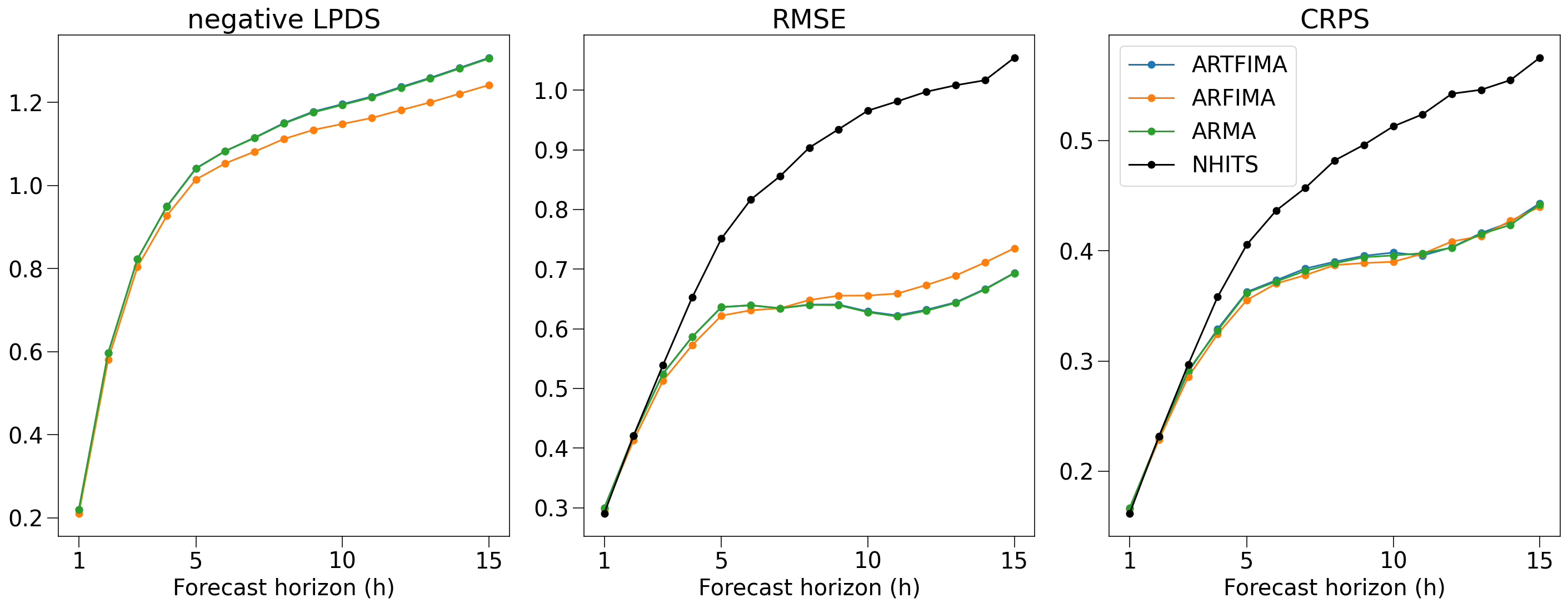}
\centering
\caption{New England electricity data: negative log-predictive density score (LPDS), root mean square error (RMSE) and the continuous rank probability score (CRPS) for all models based on $h$-step ahead forecasts for $p=2, q=1$.}
\label{fig: forecast metrics gefcom 2,1}
\end{figure}

\begin{figure}[ht]
\includegraphics[width=12cm, height=7cm, keepaspectratio]{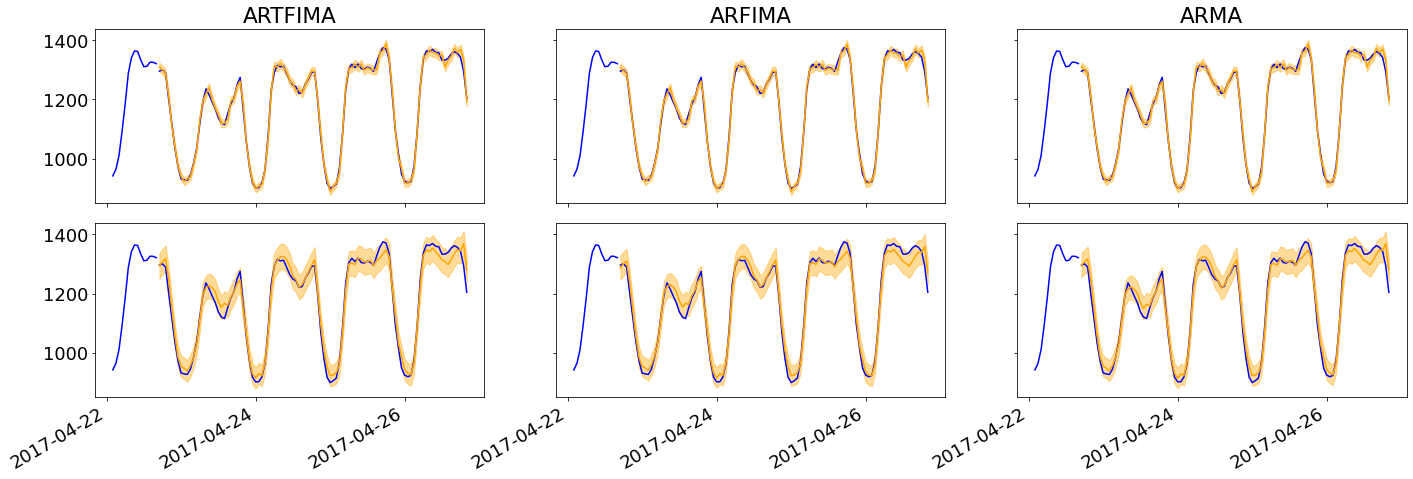}
\centering
\caption{Prediction and prediction intervals for one hour ahead ($h=1$, top panel) and ($h=6$, bottom panel) predictions for three error processes (columns). The results are shown for New England electricity demand for $p=2, q=1$ after adding back the season and trend.}
\label{fig: h_1_6 gefcom 2,1}
\end{figure}

\begin{figure}[ht]
\includegraphics[width=12cm, height=7cm, keepaspectratio]{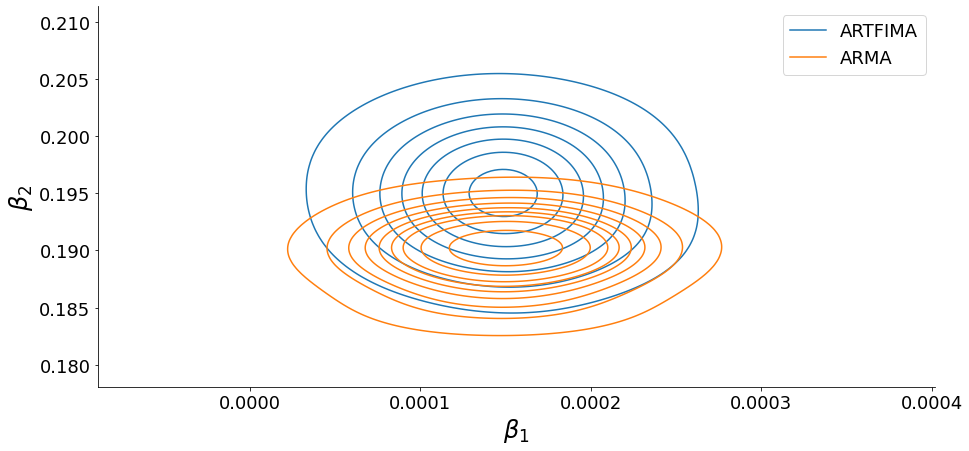}
\centering
\caption{Kernel density estimates of the posterior densities of the regression parameter $\beta$ using $10,000$ MCMC samples. The results are for the New England data with $p=2,q=1$ and two error processes, ARTFIMA and ARMA.}
\label{fig: beta_posterior 2,1}
\end{figure}

\begin{figure}[ht]
\includegraphics[width=16cm, height=7cm, keepaspectratio]{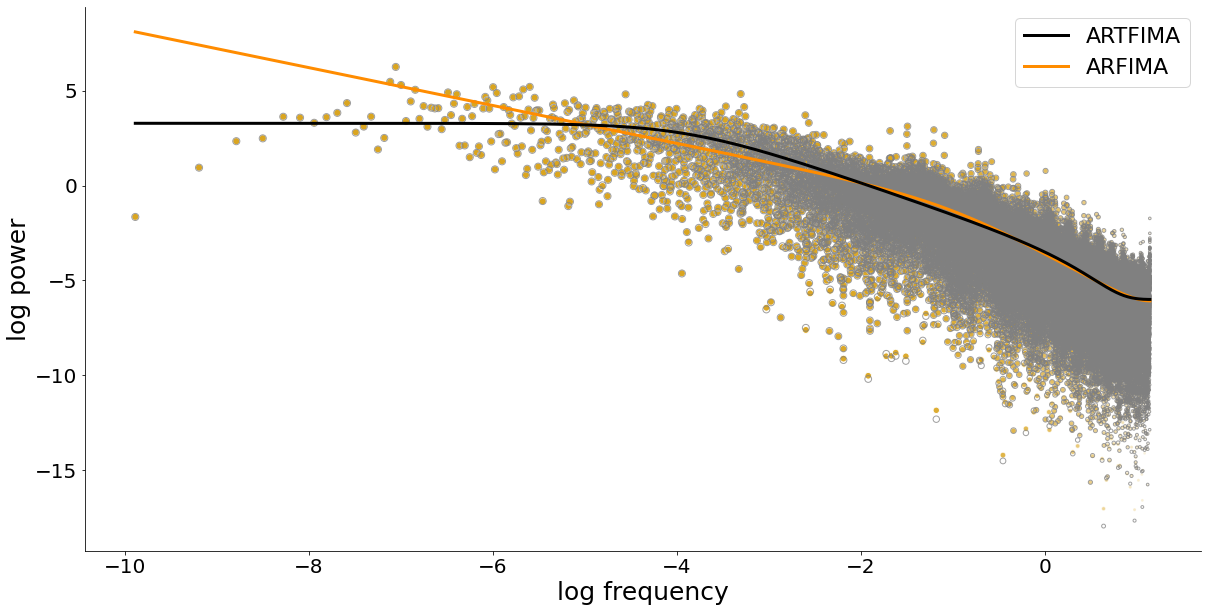}
\centering
\caption{Spectral densities at the MAP and their respective periodograms. The spectrum of DLR with ARTFIMA$(2,d,\lambda,1)$ errors with its periodogram (grey circles) and the DLR with ARFIMA$(2,d,1)$ errors with its corresponding periodogram (orange dots).}
\label{fig: loglog gefcom}
\end{figure}

Finally, Figure~\ref{fig: loglog gefcom} plots the fitted spectral densities (log-log scale) of the dynamic regression models with ARTFIMA and ARFIMA errors together with their corresponding periodogram data. Due to the dependence on $\v \beta$ (estimated $\widehat{\v \beta}$ different for the two processes), the two periodograms are different for every frequency $\omega$ but are almost identical at the lower frequencies. Consistent with the results of \cite{sabzikar2019parameter}, the ARTFIMA spectrum provides a better overall fit to the periodogram data, in particular for small frequencies.

\subsection{Example 2: Victorian electricity demand} \label{sec: vic elec}

\begin{figure}[ht]
\includegraphics[width=15.5cm, height=7cm, keepaspectratio]{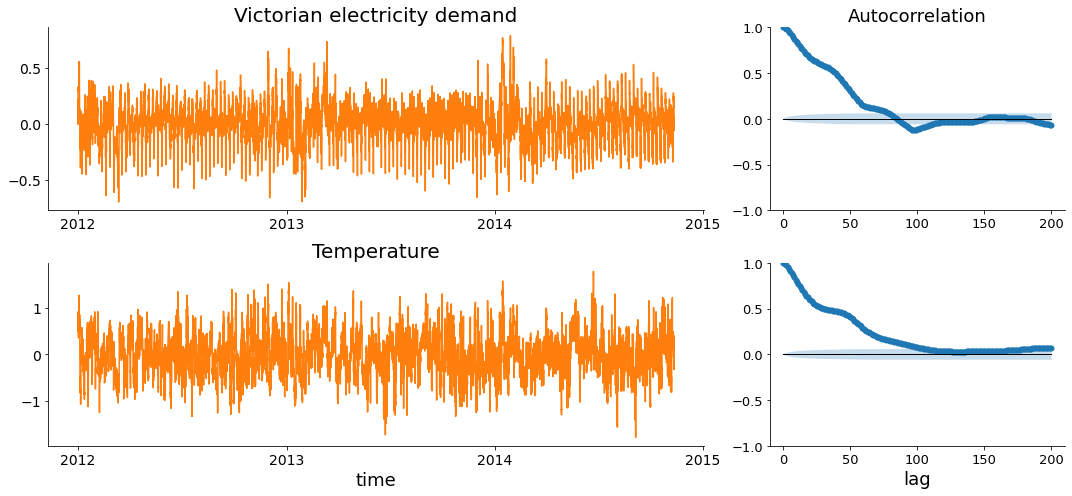} 
\centering
\caption{The left panels plot half-hourly electricity demand and temperature for Victoria, Australia after removing multi-seasonal and trend components and applying a Box-Cox transformation. The right panels are the corresponding autocorrelation plots.}
\label{fig: DLR vic elec}
\end{figure}

The second application concerns the half-hourly electricity demand for Victoria, Australia, presented in \cite{hyndman2018forecasting}. The data consists of $T=52{,}608$ observations of electricity demand (megawatts) and Melbourne's temperature (degrees Celsius) between the 1st January 2012 - 31st December 2014. The response $y_t$, Victorian electricity demand, is modelled as a linear combination of the lagged exogenous variable Melbourne temperature $x_{t-1}$. Strong multi-seasonal and trend components exist within each variable and are removed via the \texttt{mstl} function. Figure \ref{fig: DLR vic elec} depicts each variable after removing the multi-seasonal and trend components and applying a Box-Cox transformation. 

\begin{figure}[ht]
\centering
\includegraphics[width=16.5cm, height=7.5cm, keepaspectratio]{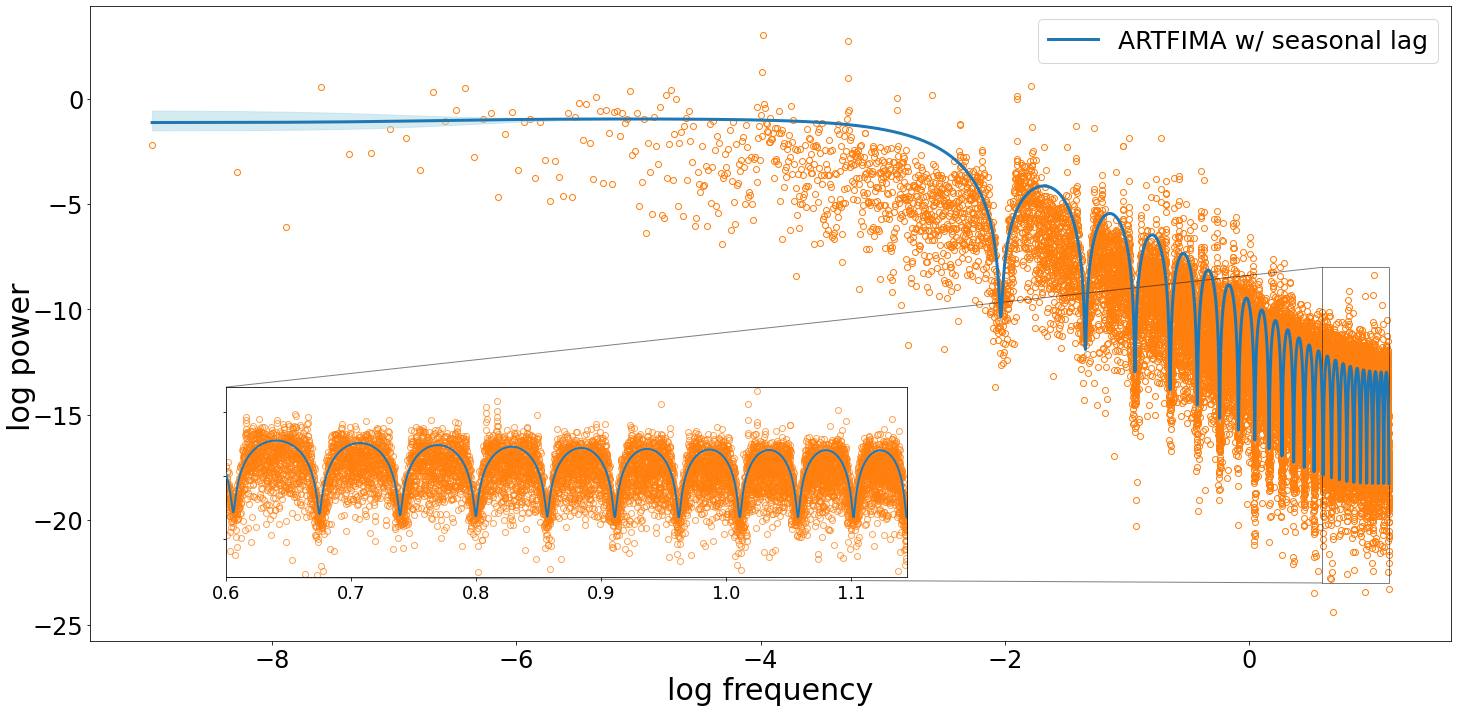}
\caption{Outer plot: Log frequency vs log power (log-log) plot of the periodogram and the spectral density function at the MAP for DLR with SARTFIMA$(2,d,\lambda, 1)(0,0,1)_{48}$ errors. Inner plot: Same image but displaying moderate frequencies on a linear scale. The 95\% credible interval of the spectral density estimated is the shaded blue region.}
\label{fig: vic elec loglog}
\end{figure}

Despite using \texttt{mstl} to remove seasonality, Figure \ref{fig: DLR vic elec} shows that all seasonality has not been removed and still exists at multiples of the 48th lag (1 day is 48 half-hours). This is also present in the periodogram, see Figure \ref{fig: vic elec loglog}, which peaks at the corresponding frequencies. For this reason, we include one seasonal MA term $Q=1$ with $s=48$ for ARMA and ARTFIMA and denote this parameter $\Psi^\star_1$ as in \eqref{eq:Standard_DLR_seasonal}. We do not include a seasonal ARFIMA model as the Gaussian likelihood is expensive to compute and our paper has overwhelming evidence that ARTFIMA captures sufficient long-range dependence while being computationally much more efficient. 

The lowest DIC for the error processes occurred for two sets of lags, $p=2,q=0$ (ARTFIMA) and $p=2, q=1$ (ARMA), which we denote as ARTFIMA$(2, d, \lambda, 0)$ and ARMA$(2, 1)$. Table \ref{tbl: vic elec models DIC} displays the DIC value and the average time for one log-likelihood evaluation for all models. For $p=2, q=0$, the lowest DIC (in bold) is the SARTFIMA model, and the worst (highest) is ARMA. ARTFIMA and ARFIMA have similar values for DIC. For the $p=2, q=1$ case, SARTFIMA obtained the lowest DIC (in bold), followed by SARMA. Again, ARTFIMA and ARFIMA produced similar DIC values. Referring to the computation times presented in Table \ref{tbl: vic elec models DIC}, significant computational speedups are gained for ARTFIMA and ARMA models, which had the shortest run time out of all models due to estimation via the Whittle likelihood. This is followed by the seasonal models, which are roughly 10 times slower than their non-seasonal counterparts. The slowest run times are observed for the ARFIMA model via the exact Gaussian log-likelihood, which is approximately 50 times slower than the non-seasonal ARMA and ARTFIMA models.

\begin{table}[ht]
\centering
\begin{tabular}{l*{6}{c}r}
\hline 
         & SARTFIMA &  ARTFIMA & ARFIMA & SARMA & ARMA \\
\hline
$p=2, q=0$      & $\bb{-341016.47}$    & $-328624.02$ & $-328612.70$ & $-338582.68$ &  $-327041.24$ \\
$p=2, q=1$        & $\bb{-341541.07}$ & $-328514.60$ & $-328587.06$ & $-338646.40$ & $-328085.01$  \\
\hline
Time (ms) & 35.1ms & 3.96ms & 203.0ms & 35.1ms & 3.79ms \\
\hline
\end{tabular}
\caption{DIC values and the average time (milliseconds) for one log-likelihood evaluation for each dynamic regression model for Victorian electricity demand.}
\label{tbl: vic elec models DIC}
\end{table}

\begin{figure}[ht]
\includegraphics[width=15cm, height=8cm, keepaspectratio]{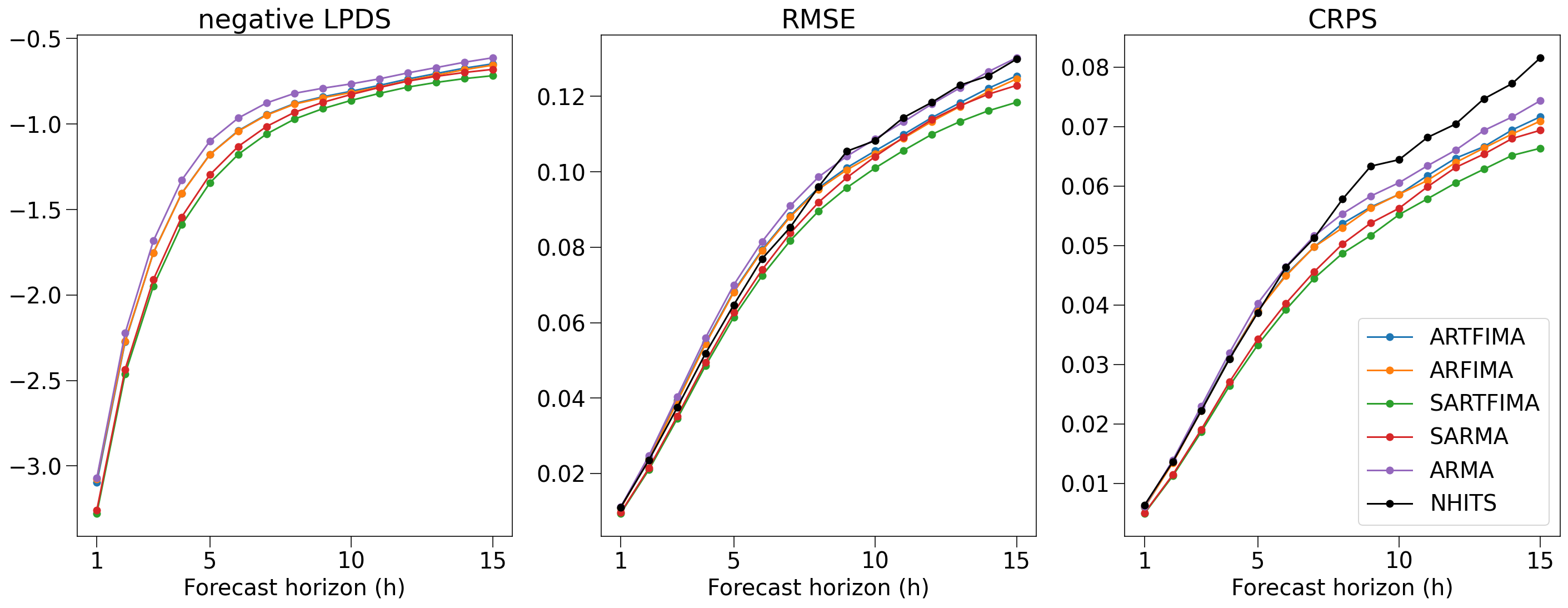}
\centering
\caption{Victorian electricity data: negative log-predictive density score (LPDS), root mean square error (RMSE) and the continuous rank probability score (CRPS) for all models based on $h$-step ahead forecasts for $p=2, q=0$.}
\label{fig: forecast metrics vic elec 2,0}
\end{figure}

Figure \ref{fig: forecast metrics vic elec 2,0} reports the forecast negative LPDS, RMSE and CRPS of the five models for the $p=2, q=0$ case for all horizons $h=1,\dots,15$. The seasonal models are better across all three metrics compared to the non-seasonal models. The seasonal ARTFIMA model has the best forecasting ability for all three metrics, closely followed by SARMA. In contrast, the ARMA model has the highest values (worst) for all three metrics among the dynamic linear regression models. Furthermore, NHITS attains RMSE values comparable to those of the ARMA model but performed worst in terms of CRPS, particularly at large forecast horizons. The ARTFIMA and ARFIMA models were virtually identical, having lower negative LPDS, RMSE and CRPS scores compared to ARMA. 

\begin{figure}[ht]
\includegraphics[width=15cm, height=8cm, keepaspectratio]{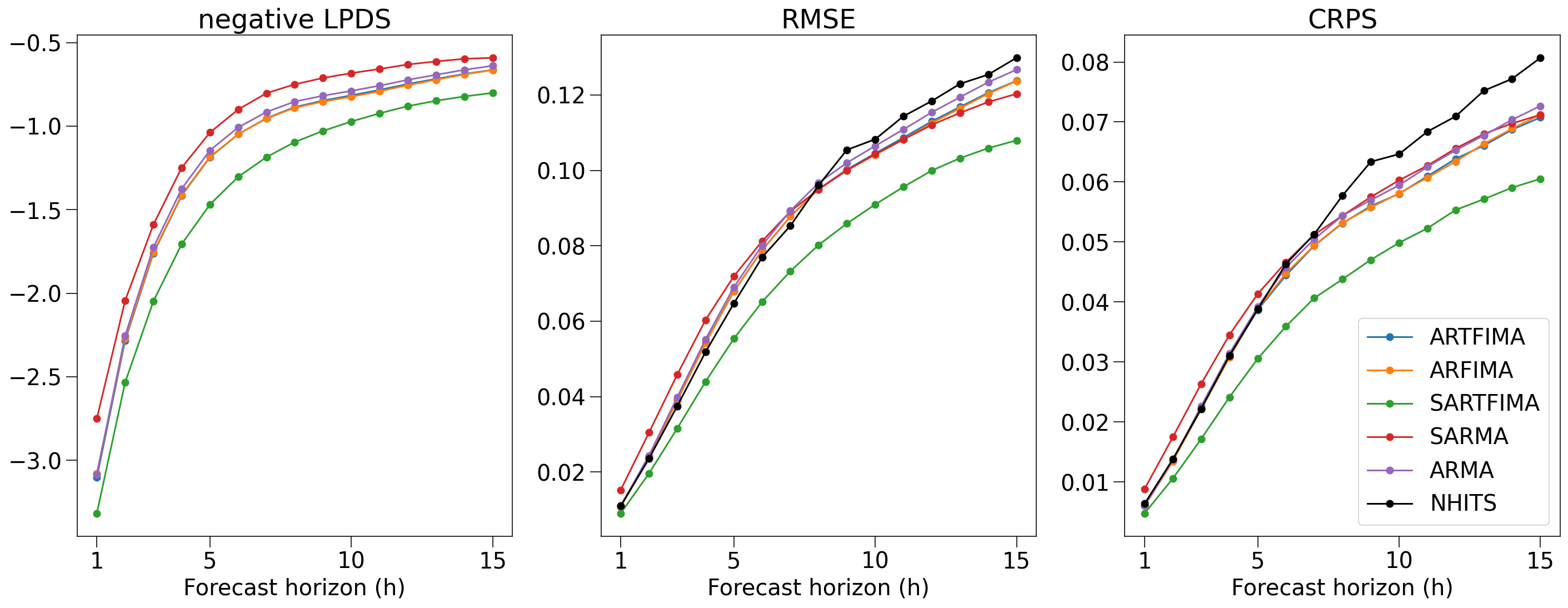}
\centering
\caption{Victorian electricity data: negative log-predictive density score (LPDS), root mean square error (RMSE) and the continuous rank probability score (CRPS) for all models based on $h$-step ahead forecasts for $p=2, q=1$.}
\label{fig: forecast metrics vic elec 2,1}
\end{figure}

For the $p=2, q=1$ case, Figure \ref{fig: forecast metrics vic elec 2,1} reports the forecast negative LPDS, RMSE and CRPS of the five models for all horizons $h=1,\dots,15$. The superiority of the SARTFIMA model is more pronounced compared to the $p=2, q=0$ case (Figure \ref{fig: forecast metrics vic elec 2,0}) across all metrics. The NHITS model is comparable to the dynamic linear regression models for short forecast horizons, but performs worst in terms of RMSE and CRPS for longer forecast horizons.

Figure \ref{fig: h_2_12 vic elec 1,2} plots the predictions and prediction intervals for the seasonal models SARTFIMA and SARMA when predicting the electricity demand one and six hours ahead, i.e.\ $h= 2,12$. While the one-hour ahead prediction is accurate for both models with seasonal ARTFIMA slightly more accurate, the seasonal ARTFIMA clearly performs better for the six-hours ahead forecast.

\begin{figure}[ht]
\includegraphics[width=12cm, height=7cm, keepaspectratio]{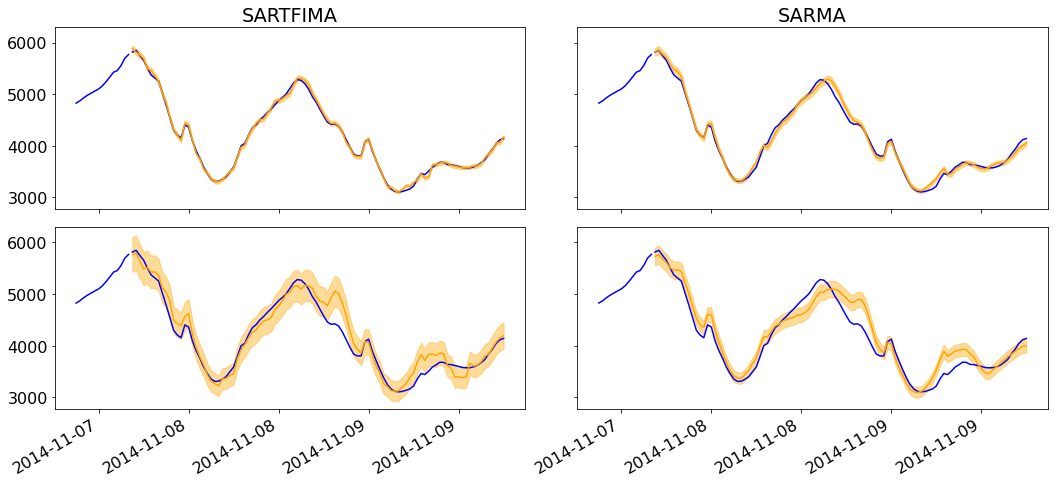}
\centering
\caption{Prediction and prediction intervals for one hour ahead ($h=2$, top panels) and six hours ahead ($h=12$, bottom panels) predictions for the two seasonal error processes (columns) for $p=2, q=1$. The results are shown for Victorian electricity data after adding back the trend and seasonal components.}
\label{fig: h_2_12 vic elec 1,2}
\end{figure}

Finally, Figure \ref{fig: vic elec loglog} further shows the appropriateness of the seasonal ARTFIMA model by displaying a log-log plot of the periodogram and spectral density at the MAP. It is apparent that the seasonal ARTFIMA model fits the lower frequencies, $\text{log}(\omega) < -4$, and also models the seasonal components, shown in the zoomed-in section of the plot, as it captures the peaks and troughs of the periodogram accurately.

\section{Conclusion and future research}

We propose a dynamic linear regression model with an ARTFIMA error process that accounts for long-range dependence. We develop a frequency domain estimation method that utilises the computationally efficient FFT. The frequency domain method uses a likelihood based on asymptotic properties of the periodogram, and we demonstrate empirically that the resulting inference matches well with time domain likelihood methods. This is expected, as the case without exogenous variables has strong theoretical results indicating that the approximation error vanishes asymptotically.

We show empirically that the well-known phenomena of the spectral density of the ARFIMA model not fitting the periodogram data well for small frequencies carries over to the dynamic linear regression case. As such, we argue that dynamic linear regression models with ARFIMA errors are unsuitable for frequency domain methods, and their estimation needs time domain methods which are computationally expensive. In contrast, we show that the dynamic linear regression models with ARTFIMA errors fit the low-frequency spectrum well and are thus a computationally fast alternative to model long-range dependence.

Our proposed models demonstrate superior forecasting performance when the data exhibits complex long-range dependence and perform on par with the ARMA error process for simpler data. This is expected as our model nests the dynamic linear regression model with ARMA errors. We do not observe any signs of overfitting when using an ARTFIMA error process instead of an ARMA for the simpler dataset. In addition, we also observe that the posterior distribution of the regression parameter is significantly affected by not accounting for long-range dependence in the error process. We also compare our proposed model to a neural-network-based forecasting approach. In our applications, the neural network is consistently outperformed by the proposed model across most forecast horizons and evaluation metrics. These results indicate that, after removing seasonality and trend, a parsimonious semi-long memory model can deliver substantially better forecasting performance than a flexible deep-learning based alternative.  In summary, we believe dynamic linear regression models with ARTFIMA errors are robust and should be routinely included in a forecaster's toolkit. 

Future research will extend our approach to modelling a multivariate response as in \cite{villani2022spectral}. For small to moderate datasets, the Whittle log-likelihood is known to be severely biased. Exploring recent advances such as the debiased Whittle likelihood \citep{sykulski2019debiased} for dynamic linear regression models can potentially enable our method to be used for small to moderately large datasets. Finally, the structure of our model can utilise subsampling MCMC methods \citep{quiroz2019speeding, quiroz2021block} in the spectral domain \citep{salomone2020spectral}.

\bibliographystyle{apalike}
\bibliography{ref}

\end{document}